\documentclass[prx,amsmath,amssymb, notitlepage, twocolumn,
nofootinbib,
superscriptaddress,
longbibliography
]{revtex4-1}

\usepackage{amsmath}
\usepackage{tabularx,graphicx}
\usepackage{epstopdf}
\usepackage{graphicx}
\usepackage{latexsym}
\usepackage{amssymb}
\usepackage{amsmath}
\usepackage{color}
\usepackage{psfrag}
\usepackage{bbm}
\usepackage{bm}
\usepackage{titlesec}
\usepackage{dsfont}
\usepackage{feynmp}
\usepackage{slashed}
\usepackage{multirow}
\newcommand{\mc}[1]{\mathcal{#1}}

\newcommand{\beq}{\begin{eqnarray}}
\newcommand{\eeq}{\end{eqnarray}}

\newcommand{\bsp}{\begin{split}}
\newcommand{\esp}{\end{split}}

\usepackage{color}
\definecolor{darkblue}{rgb}{0.,0.,0.4}
\definecolor{darkred}{rgb}{0.5,0.,0.}
\definecolor{BlueViolet}{RGB}{138,43,226}
\definecolor{SkyBlue}{RGB}{30,144,255}
\definecolor{DarkGreen}{RGB}{0,100,0}
\usepackage[pdftex,colorlinks=true,linkcolor=darkblue,citecolor=blue,urlcolor=darkred]{hyperref}
\usepackage[normalem]{ulem}

\newcommand{\moire} {moir{\' e} }

\renewcommand{\vec}[1]{\bm{#1}}

\begin{document}

\title{Band Structure of Twisted Bilayer Graphene: Emergent Symmetries, Commensurate Approximants and Wannier Obstructions}
\author{Liujun Zou}
\affiliation{Department of Physics, Harvard University,
Cambridge, MA 02138, USA}
\affiliation{Department of Physics, Massachusetts Institute of Technology,
Cambridge, MA 02139, USA}
\author{Hoi Chun Po }
\affiliation{Department of Physics, Harvard University,
Cambridge, MA 02138, USA}
\author{Ashvin Vishwanath}
\affiliation{Department of Physics, Harvard University,
Cambridge, MA 02138, USA}
\author{T. Senthil}
\affiliation{Department of Physics, Massachusetts Institute of Technology,
Cambridge, MA 02139, USA}

\begin{abstract}

A remarkable feature of the band structure of bilayer graphene at small twist angle is the appearance of isolated bands near neutrality, whose bandwidth can be reduced at certain magic angles (eg. $\theta\sim 1.05^\circ$).  In this regime, correlated insulating states and superconductivity have been experimentally observed. A microscopic description of these phenomena requires an understanding of universal aspects of the band structure, which we discuss here. First, we point out the importance of emergent symmetries, such as valley conservation, which are excellent symmetries in the limit of small twist angles and dictate qualitative features of the band structure. These have sometimes been overlooked when discussing commensurate approximants to the band structure, which we also review here, and solidify their connection with the continuum theory which incorporates all emergent symmetries. Finally, we discuss obstructions to writing down tight-binding models of just the isolated bands, and in particular a new symmetry based diagnostic of these obstructions, as well as relations to band topology and strategies for resolving the obstruction. Especially, we construct a four-band model where the two lower isolated bands realize all identified Wannier obstructions of the single-valley nearly flat bands of twisted bilayer graphene.

\end{abstract}
\maketitle


\section{Introduction}

Following the recent discovery of correlated insulating and superconducting states in ``magic angle'' twisted bilayer graphene (TBG) \cite{Cao2018a, Cao2018}, there has been considerable theoretical activity seeking to define an appropriate low energy model \cite{Volovik2018, Xu2018, NoahLiang, Mar2018, Roy2018, Guo2018,  Baskaran2018, Padhi2018, Irkhin2018, Dodaro2018, Huang2018, Zhang2018, Ray2018, Liu2018, XuLawLee2018, Oskar, Rademaker2018, Isobe2018, KoshinoLiang, Wu2018, Pizarro2018, Peltonen2018, You2018, WuXu2018, Pal2018, Ochi2018, Fidrysiak2018, Thomson2018, Guinea2018}.  In our previous work \cite{Mar2018}, we argued in favor of a honeycomb lattice description despite the concentration of charge density at the sites of a triangular lattice. Similar conclusions were also reached in Refs. \onlinecite{NoahLiang, Oskar, KoshinoLiang}. Further we argued that the system has a number of excellent symmetries (even if all of them are not exact) which together present an obstruction to constructing well-localized Wannier functions. Specifically any  such Wannier functions, even centered on honeycomb sites, will not transform into themselves at each site under the symmetry operations, i.e., some symmetry actions become unnatural. By sacrificing on-site action for a valley $U(1)$ symmetry, we showed  explicitly that Wannier functions can indeed be constructed, and showed how to correctly identify the ``non-local'' valley charge operator.

In this paper we discuss several new aspects of this obstruction. In Ref. \onlinecite{Mar2018}  we identified two seemingly distinct obstructions - one tied to the same chirality of the two Dirac nodes in the spectrum at each valley, and the other tied to the representations of a ``mirror''\footnote{This actually is a 180 degree rotation in $3d$ about an axis that lies parallel to the two graphene layers.} symmetry if present. Here we show that these two obstructions are, in fact, the same. The mirror symmetry, when present, along with the Dirac points implies the same chirality Dirac nodes in the band structure.
This result leads to a powerfully simple way to identify the chirality obstruction by simply examining the mirror eigenvalues at the $\Gamma$ point of the \moire Brillouin zone (mBZ).  We also give a  physical discussion of the chirality obstruction by considering the effects of breaking a $C_2$ (180 degree rotation in the 2D plane) symmetry.  Along the way we clarify many other confusing aspects of the band aspects of TBG, and in particular the nature and role of its global symmetries.

At a generic incommensurate twist angle, the twisted bilayer structure has very little symmetry. The only exact symmetries are $U(1)$ charge conservation, time reversal $\mc{T}$, and $SU(2)$ spin rotation (ignoring the weak spin-orbit coupling).  In particular, it is not even translation invariant. It is natural then to wonder if there are well defined bands at all in the first place. This has led some authors to restrict attention to special commensurate structures with large periods as a clear theoretical system to discuss the band structure. Experimentally well-defined band gaps induced by the moire superlattice are seen. In near magic angle samples the gaps are estimated to be $\approx 35 meV$ which are much bigger than the expected band width of the nearly flat low energy bands.

Theoretically at small twist angles, there is a well known ``continuum theory'' description which yields well-defined band structures for all twist angles including incommensurate ones \cite{Neto2007, Bistritzer2011}. The continuum theory reveals many universal features of the band structure, such as the existence of Dirac crossings between valence and conduction bands within each valley of the underlying graphene layers. These features have been benchmarked against tight-binding calculations on commensurate structures \cite{Shallcross, Morell2010, Mayou2010, Jung2014}. They are also nicely consistent with experiments at twist angles larger than the magic angles (where correlation effects are expected to be weaker, and band theory predictions can be reasonably compared with experiment). In particular, Cao {\it et al} showed that at a twist angle of about 1.8 degrees the Landau fan structure near charge neutrality is exactly what is expected from the Dirac points predicted by the continuum theory \cite{PabloPRL}.  Despite its success for qualitative universal aspects, quantitatively the continuum theory yields a very small value compared to experiments for the gap separating the nearly flat bands from other bands. This discrepancy  is  believed to be reduced once effects of lattice relaxation and electron interactions are included. Formally these additional effects can be included phenomenologically in the continuum model by modifying its parameters away from those estimated microscopically \cite{Mar2018, KoshinoLiang}.

Apart from translational symmetry, the approximations involved in the continuum theory build in a number of other point group symmetries which are not fully present in any commensurate structure.  These include a $C_6$ rotation symmetry, and a valley $U_v(1)$ associated with separate conservation of electrons associated with each valley. This symmetry structure of the continuum theory is essential in protecting the Dirac points of valley filtered bands. Specifically, on top of the valley $U_v(1)$ (needed to define separate bands within each valley)  a $C_2 {\cal T} = C_6^3 {\cal T}$ symmetry is able to protect the Dirac points from acquiring a gap. Even restricting to commensurate structures with translation symmetry, $U_v(1)$, and maybe even $C_2\mc{T}$, are not both exact microscopic symmetries.

In the older literature it was appreciated that at small twist angles  the extra symmetries of the continuum theory are excellent  approximations \cite{Shallcross, Morell2010, Mayou2010, Castro-Neto2012}. Furthermore it was understood that there is essentially no difference between incommensurate and commensurate structures, or between distinct commensurate structures with different exact microscopic symmetries. These issues have re-emerged in  recent discussions of TBG, and have led to some confusion.  We therefore carefully review and collect together some pertinent facts about different commensurate structures, their relationship to the continuum theory, and the
implications for a description of small angle (possibly incommensurate) TBG.

The most fundamental aspect of our previous discussion of the  band structure is the existence of an obstruction to constructing well localized Wannier functions transforming naturally under all symmetry operations.
The obstruction relies strongly on the presence of symmetries that are not exact microscopic symmetries. Why then should we worry about it? Let us therefore review the tight logic that forces us to confront it.
As reviewed above, it is a robust feature of both theory and experiment that to excellent accuracy there is a good valley $U_v(1)$ symmetry and that within each valley there are  Dirac band crossings (down to energy scales currently accessible in experiments).   The robustness of the Dirac crossings within each valley suggests that it is a symmetry protected feature of the band structure. The natural protecting symmetry then is $C_2{\cal T}$ as is seen explicitly in the continuum theory. For a general small-angle, incommensurate TBG structure, the $C_2$  symmetry---like translations itself---is not an exact symmetry, but it must be excellent enough to give the Dirac cones. This then forces us to study systems which have translations, valley $U_v(1)$, and $C_6$ as good symmetries. However the implementation of all these symmetries in the band structure leads to a Wannier obstruction.

Suppose we took the opposite logic, and study a commensurate structure with, say, an exact superlattice translation, and an exact  $D_3$ symmetry, and ignored all emergent approximate symmetries. This is done by Ref. \onlinecite{Oskar}.   It is then indeed possible to follow the usual procedure and construct Wannier  functions for the low energy nearly flat bands that respect the assumed exact symmetries.  The resulting tight-binding model is then shown to have Dirac points \cite{Oskar}. This procedure is certainly not mathematically wrong. How then should we reconcile it with our claims on Wannier obstructions? The point is that the assumed exact symmetries in Ref. \onlinecite{Oskar} are not enough to protect the Dirac points which must therefore be viewed as fine-tuned features of the constructed tight-binding model.  A generic perturbation allowed by the assumed exact symmetries will gap out the Dirac points. The robustness of the Dirac points observed in experiments and band structure then  forces us to include extra symmetries which must be emergent in the limit of small twist angles. We should then ask how these extra symmetries are implemented on the Wannier functions, and we are again led to the obstruction with implementing them as ``on-site'' symmetries. Similar remarks apply also to treatments that start with the continuum model and build Wannier functions by ignoring one or other of the symmetries of the model \cite{KoshinoLiang}. It is then important to know how the ignored symmetry is actually implemented in the resulting tight-binding model. We described this for the valley $U_v(1)$ in Ref. \onlinecite{Mar2018}. In contrast, Ref. \onlinecite{KoshinoLiang} ignored the $C_2$ symmetry. It remains to be seen exactly how the $C_2$ present in the continuum model is then represented non-locally in their tight-binding model.

We also clarify an apparent discrepancy between our work \cite{Mar2018} and  Refs. \onlinecite{NoahLiang, Oskar, KoshinoLiang} which is less subtle, concerning the realization of $C_3$ rotation symmetry.  Specifically, we show that this difference can be traced to a difference in the definition of the rotation center for the $C_3$ operation.

In the recent theoretical literature, some authors have proposed describing the system by a triangular lattice tight-binding model with 2 spins and 2 orbitals (presumably corresponding to the 2 valleys) per lattice site \cite{Xu2018, Guo2018, Dodaro2018, Fidrysiak2018}. This is motivated by the known concentration of the charge density at the sites of a triangular lattice. However, this triangular description is inconsistent with the existence of Dirac points together with the known symmetry representations at high symmetry points in the band structure.

We begin the rest of the paper by carefully reviewing distinct commensurate structures and their symmetries in Sec. \ref{sec:review} and Sec. \ref{sec:comsym}. Next we review  the continuum model and show how it captures universal aspects of the band structure, and further agrees with the results of existing calculations on commensurate structures in Sec. \ref{sec:cont}. Along the way, in Sec. \ref{sec: remark_magic_angle} we will make some remarks on the notion of magic angle. We then describe our new results on the obstruction in Sec. \ref{sec: relation_obstruction}, and discuss how to resolve the obstruction in Sec. \ref{sec: resolve}. Finally, we summarize the results in Sec. \ref{sec: discussion}.

\section{Commensurate structures: Types I and II structures with D3 or  D6 symmetry
\label{sec:review}}
We start by reviewing some purely geometric aspects of commensurate TBG.
The content of this section has all been discussed in existing literature \cite{Mele2010,Shallcross,Castro-Neto2012}; here, we simply review these results in an attempt to clarify some potential confusions which arose in light of the recent interest of TBG.

Imagine constructing a TBG system as follows: first, we stack the two monolayers on top of each other in a site-by-site manner (i.e., an ``AA-stacked'' bilayer); second,  we rotate the top layer counter-clockwise by a twist angle $\theta$ about a chosen point in space.
For generic $\theta$, the resulting crystal structure exhibits \moire pattern but does not have {\it exact} lattice translation symmetries.
We say the structure is commensurate when the twist angle is special such that some (moir{\'e}) translation symmetries are retained.
This happens when the twist angle takes the following form:
\beq \label{eq: commensurate angles}
\cos\theta(m, r)=\frac{3m^2+3mr+r^2/2}{3m^2+3mr+r^2},
\eeq
where $m$ and $r$ are coprime positive integers (we follow the notations in Ref.\ \onlinecite{Castro-Neto2012} here). Here, we restrict the twist angle to be in $(0,\pi/3)$, which is sufficient for our discussion as any other twist angle can be related to one in this range using the symmetries of the system.
The twist angles in Eq.\ \eqref{eq: commensurate angles} are called commensurate angles.

One important aspect that deserves immediate clarification is that the commensuration condition of the bilayer depends only on the twist angle, but not the twisting center (which is not specified above). That is, as long as the twist angle has the form in Eq.\ \eqref{eq: commensurate angles}, the bilayer has exact translation symmetries even when the twisting center is chosen to be a generic point such that {\it no where} in the lattice are two carbon atoms perfectly aligned.
It is worth emphasizing that the \moire lattice vectors of a commensurate structure are determined {\it solely} by the twist angle $\theta$.
Consequently, the \moire lattice constant $L(m,r)$, as well as the mapping of momenta between the microscopic Brillouin zone (BZ) and the \moire BZ, have no bearing on the choice of the twisting center. Rather, the twisting center determines the {\it exact} symmetries of the commensurate lattice, as we will review later.

Let us now consider such properties.
To fix conventions, let $\vec a_1 = a (1,0)$ and $\vec a_2 = a (-1/2, \sqrt{3}/2)$  be the primitive lattice vectors of the unrotated layer, and $\vec t_1$, $\vec t_2$ be the \moire lattice vectors.
The commensurate angles are further divided into two types according to how $\vec t_{1,2}$ and $\vec a_{1,2}$ are related to each other\footnote{The following expressions are slightly different from those in Ref.\ \onlinecite{Castro-Neto2012} due to our different choice of $\vec a_1$.
}:\\
~\\
\noindent {\it Type I}: if ${\rm gcd}(r, 3)=1$, then
\beq
\left(
\begin{array}{c}
	\vec t_1\\
	\vec t_2
\end{array}
\right)
=
\left(
\begin{array}{cc}
m & 2m+r\\
-(m+r) & m
\end{array}
\right)
\left(
\begin{array}{c}
	\vec a_1\\
	\vec a_2
\end{array}
\right).
\label{eq:typeI}
\eeq
~\\
\noindent {\it Type II}: if ${\rm gcd}(r, 3)=3$, then
\beq
\left(
\begin{array}{c}
	\vec t_1\\
	\vec t_2
\end{array}
\right)
=
\left(
\begin{array}{cc}
	m+\frac{r}{3} & m+ \frac{2r}{3}\\
	-\frac{r}{3} & m+\frac{r}{3}
\end{array}
\right)
\left(
\begin{array}{c}
	\vec a_1\\
	\vec a_2
\end{array}
\right).
\label{eq:typeII}
\eeq
where ${\rm gcd}(m, n)$ denotes the greatest common divisor of the integers $m$ and $n$. It follows that the \moire lattice constant is
\beq
L(m ,r)=a\sqrt{\frac{3m^2+3mr+r^2}{{\rm gcd}(r, 3)}}.
\label{eq:Lmr}
\eeq
Note that this formula applies to both types I and II structures.

\begin{figure}[h]
\begin{center}
{\includegraphics[width=0.45 \textwidth]{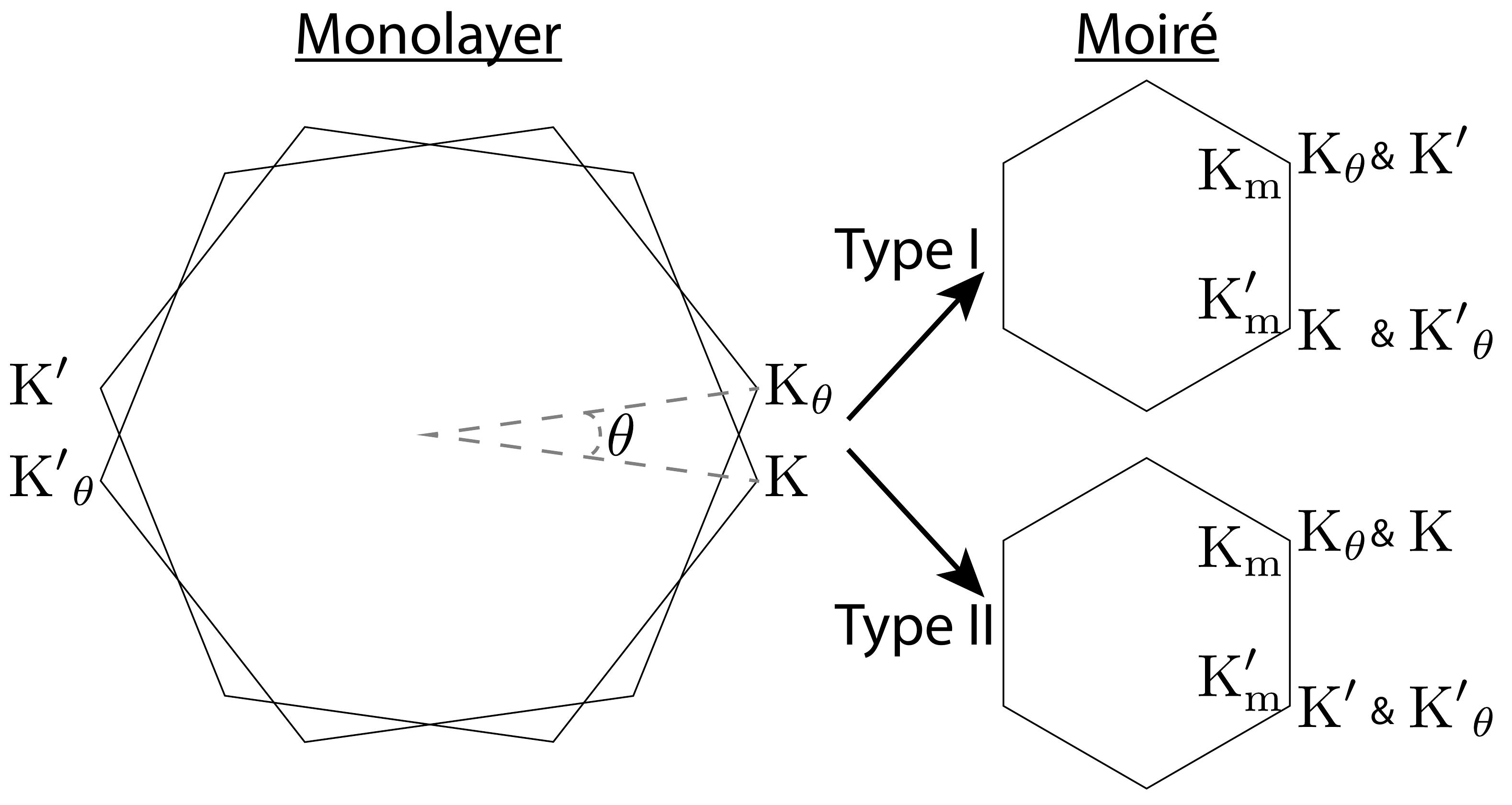}}
\caption{ Pattern of momentum mapping for the two different types of twist angles. Here, K, K$_{\theta}$ and K$_{\rm m}$ respectively denote the K point of the unrotated, rotated, and \moire Brillouin zones. K', K'$_{\theta}$, and K'$_{\rm m}$ denote their respective time-reversal partners.
\label{fig:folding}
 }
\end{center}
\end{figure}

The described relations between $\vec t_{1,2}$ and $\vec a_{1,2}$ fix the corresponding relations between the \moire and monolayer reciprocal lattice vectors. In particular, this establishes a folding from the monolayer BZ to the \moire BZ. The different forms in Eqs.\ \eqref{eq:typeI} and \eqref{eq:typeII} naturally lead to different folding patterns for type I and II structures.
Let $\vec K$ be the momentum of the K point of the unrotated layer, $\vec K'\equiv -\vec K$ be that of the K' point, and $\vec K_\theta \equiv R_{\theta} \vec K$ be the corresponding momentum in the rotated layer ($R_{\theta}$ being the counter-clockwise rotation matrix by angle $\theta$).
One can then verify that, for both types I and II structures, each of the momenta $\pm \vec K$ and $\pm \vec K_\theta$ is folded to a \moire K point. From time-reversal (TR) symmetry, we must have $\vec K$ and $- \vec K$ mapping to different \moire K points, and similarly for $\vec K_\theta$ and $-\vec K_\theta$. So, to determine the pairing pattern, one simply checks if $\vec K \pm \vec K_\theta$ is a \moire lattice vector. The resulting pattern is shown in Fig.\ \ref{fig:folding}.


%

We now specialize to small twist angles, say $\theta \sim 1^\circ$, relevant for the recent experiments
\cite{Cao2018,Cao2018a}.
A first observation is that $|\vec K - \vec K_\theta| \ll |\vec K' - \vec K_\theta|$ for $0<\theta\ll1$, with the former of $\mathcal O(\theta/a)$ and the latter of $\mathcal O(1/a)$. Such separation of scales has important physical consequences.
As the \moire potential is slowly varying in space with a typical length scale set by $\sim a/\theta$, the coupling between Bloch states with large momentum difference is ineffective even if it is symmetry-allowed.
Therefore, for type I structures, although K$_\theta$ and K' are folded to the same \moire momentum, a direct coupling between the corresponding Dirac points is suppressed in the small angle limit.
We refer to such effective decoupling of degrees of freedom as a ``valley charge conservation'', where the states in the vicinity of the microscopic $K$ and $K_\theta$  points are grouped into a valley, and their TR partners into the other.

The case for type II, however, is apparently different. As can be seen from Fig.\ \ref{fig:folding}, K$_\theta$ and K, which belong to the same valley, are now  folded to the same \moire K point. In principle, there is neither a symmetry nor energetics reason to forbid the direct coupling between their associated Dirac points.
This conclusion, however, is drawn using only the {\it exact} symmetries of the commensurate lattice, and ignores the presence of approximate symmetries in the problem. As we will see, such small-angle type-II structures will  feature approximate translation symmetries that suppress the direct coupling between the Dirac points in the same valley.

To this end, let us first introduce another length scale $L' (\theta) \equiv a/(2\sin\frac{\theta}{2})$, which is known to determine the experimentally observed \moire lattice constant from scanning tunneling microscopy \cite{STMNatPhy, STMPRL,  Crommie2015, Kim2017}.
Generally, $L$ and $L'$ do not coincide:
\beq
\frac{L(m,r)}{L' (m,r)} = \frac{r}{\sqrt{{\rm gcd}(r,3)} } \geq 1,
\eeq
and so $L = L'$ if and only if $r=1$.
It is, therefore, tempting to conclude that only the $r=1$ type-I commensurate approximants are relevant for these experiments. However, this is a misconception.

First, while Eq.\ \eqref{eq:Lmr} gives the lattice constant corresponding to exact translation symmetries of the commensurate lattice, in practice the small-angle structures are known to have approximate translation symmetries with the pitch $L'(\theta)$, even if $\theta$ does not belong to the $r=1$ type-I series \cite{Castro-Neto2012}. On the electronic states, such approximate symmetries lead to a suppression of certain coupling between the Bloch states \cite{Castro-Neto2012}. In other words, to infer the {\it exact} lattice constant from experiments, one must achieve energy resolution smaller than these suppressed coupling strengths. Absent such resolution, the experimental lattice constant at small twist angles would be $L'$, even if the device was (magically) an exact type-II commensurate lattice with, say, $r=3$.
To properly account for the physics observed at the experimental energy scale, one should incorporate such approximate symmetries into the analysis. The continuum theory \cite{Neto2007, Bistritzer2011} provides a powerful avenue to do this, as we will discuss in Sec.\ \ref{sec:cont}.

Second, even if one chooses to focus exclusively on the exact geometrical properties of commensurate TBG with the lattice constant $L'$, it is still wrong to focus only on the $r=1$ structures.
Recall that $\theta(m,r) \in (0,\pi/3)$ in Eq.\ \eqref{eq: commensurate angles}. As a graphene monolayer is invariant under a $\pi/3$-rotation, $\theta = \pi/3 - \phi$ for some $0<\phi  \ll 1$ also corresponds to a system with a small {\it physical} twist angle in the opposite sense\footnote{Clearly, a $-\phi$ rotation of the top layer is the same as a $\phi$ rotation of the bottom layer.
Also, we reiterate that, for the current discussion, we do not fix the twisting center; to reconcile a $-\phi$ structure with one with $\pi/3 - \phi$, one might need to choose different twisting centers.
}. For such systems, the physically relevant length scale is $L'(\phi)$, and indeed one can find structures where $L'(\phi)$ is identical to the exact lattice constant $L(\pi/3- \phi)$, i.e., such structures are equally good candidates as commensurate approximants to the experimental systems, although they do not belong to the $r=1$ series.
We warn that, for these ``conjugate'' structures the assignment of microscopic Dirac points into different valleys do not conform with the preceding discussion, which assumed $\theta \ll 1$ (rather, we have $\phi = \pi/3 - \theta \ll 1$ here).
More details and clarification on these ``conjugate'' structures are provided in Appendix \ref{sec:conj}. For simplicity, in the following we will ignore such conjugate structures, and always assume $\theta \ll 1$.

\begin{figure*}
{\includegraphics[width=0.95 \textwidth]{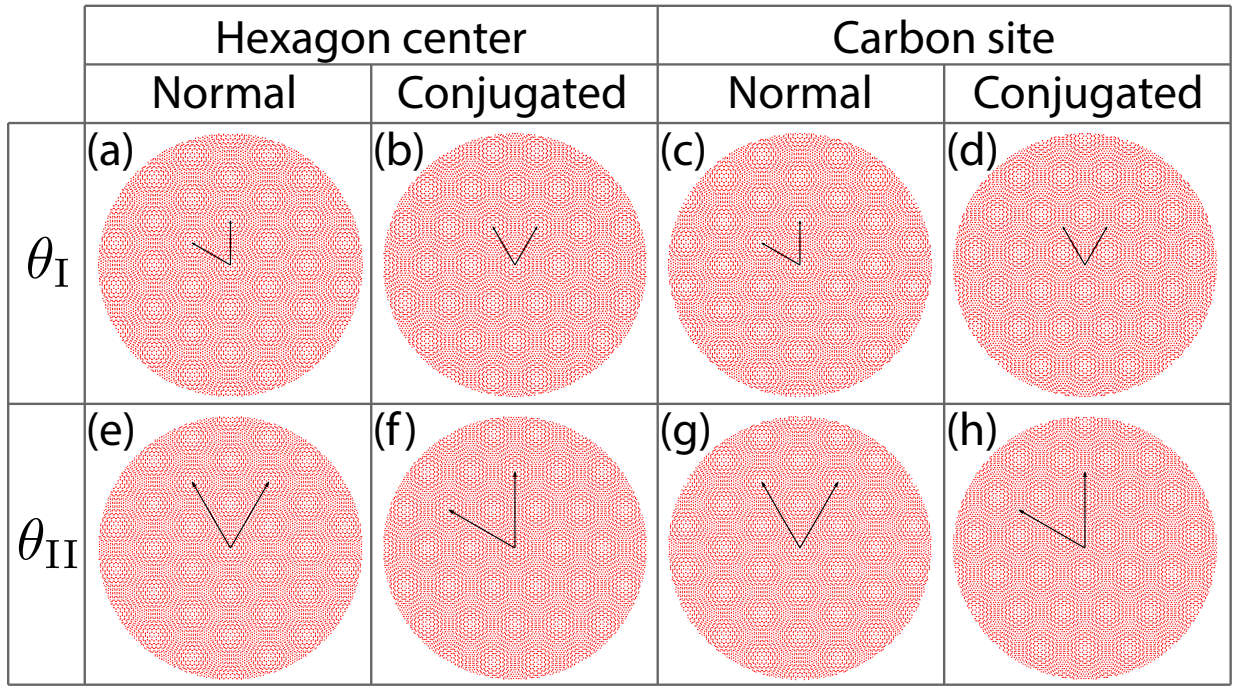}}
\caption{Examples of commensurate lattices. Here, we twist the top layer counter-clockwise  by $\theta/2$ and the bottom by $-\theta/2$.
The radius of the disk in each panel is chosen to be $3 a / (2 \sin (\theta/2))$.
Two relatively large twist angles $\theta_{\rm I} \equiv \theta(5,1)\sim 6.0^\circ$ and $\theta_{\rm II} \equiv  \theta(16,3)\simeq 5.7^\circ$ are considered. If smaller twist angles were used, the differences between the lattices would be even harder to discern.
As indicated by the subscripts, they are respectively of types I and II, which lead to different primitive lattice vectors (black arrows). The twisting centers are chosen to be either an aligned hexagon center or an aligned carbon site, and we consider both a normal twist by $\theta$, as well as a ``conjugated'' twist by $\pi/3 - \theta$. Note that conjugation changes the angle type (Appendix \ref{sec:conj}), so panels (a,c,f,h) are type I lattices whereas (b,d,e,g) are of type II. Although the {\it exact} lattice constant for lattices (e--h) is larger, one can see visually that their approximate lattice constant is the same as that of lattices (a--d), which is given by $L'(\theta) = a/(2 \sin(\theta/2))$. Except for lattices (c) and (h), the {\it exact} point group of all the lattices shown is $D_6$.
\label{fig:commensurate}
 }
\end{figure*}

Having clarified the relationship between the physical \moire pitch and the microscopic {\it exact} lattice constant of the commensurate approximants, we now move on to the point-group symmetries of the commensurate lattices.
As we have emphasized, in the discussion so far we have made no assumption on the choice of the twisting center. In particular, one could have chosen the twisting center to be a generic point in space, such that only translations remain as exact spatial symmetries of the resulting commensurate structure. Alternatively, if we choose the twisting center to be a hexagon center, the resulting TBG structure will inherent the six-fold rotation symmetry $C_6$ of the underlying monolayers. In other words, the choice of twisting center determines the {\it exact} point-group symmetries of the TBG.

Starting with an AA-stacked bilayer, there are two main choices of twisting center that lead to a high-symmetry structures: one either twists about a common hexagon center, or about a common honeycomb site (i.e., about a carbon site)\footnote{
In fact, as discussed in Appendix \ref{sec:conj}, for a certain class of angles the two choices are equivalent.
}.
Let us first consider twisting about a hexagon center. Aside from the mentioned $C_6$ rotation, there is an additional two-fold rotation about an axis running parallel to the 2D planes, i.e., one which exchanges the two layers of graphene \cite{Mar2018}. Note that, this two-fold rotation is distinct from $C_2 \equiv C_6^3$, with the latter leaving the monolayers individually invariant. To avoid confusion, we will refer to the layer-exchanging rotation as a ``mirror'' symmetry $M_y$, which is appropriate when one views the system as strictly two-dimensional. $C_6$ and $M_y$ together generate the point group $D_6$, i.e., all commensurate structures with a hexagon center chosen as the twisting center will have exact spatial symmetries described by the wallpaper group 17 ($p6mm$)\cite{ITC}.

On the other hand, if the twisting center is a common honeycomb site, the system is only invariant under a $C_3$  {\it but not} a $C_6$ rotation about the twisting center. Correspondingly, the point group of the twisting center is reduced to $D_3$, which differs from $D_6$ only by the absence of $C_6$. While one might be led to conclude that the resulting system is described by wallpaper group 15 ($p31m$)\cite{ITC}, this is only true for type I structures. As was pointed out in Ref.\ \onlinecite{Mele2010}, with this choice of twisting center the type II structures have a higher degree of symmetries, and one can show that there will always be one aligned hexagon center, with a $D_6$ symmetry, in each unit cell. Because of that, the correct wallpaper group becomes 17 ($p6mm$).

We note in passing that, in Ref.\ \onlinecite{Mele2010}, types I and II structures generated by twisting about a common honeycomb site are respectively called ``SE-odd'' and ``SE-even,'' where SE stands for ``sublattice exchange.'' Ref.\ \onlinecite{Mele2010} further pointed out that SE-odd and even structures are expected to display different generic low-energy behavior at charge neutrality, but the analysis there is relevant for large twist angles close to $\pi/6$, and does not apply to the small-angle regime where the effective valley charge conservation comes into play.

In closing, we make a small remark on terminology. For small {\it physical} twist angle $\theta$, the resulting \moire pattern is known to exhibit regions that are locally close to ``AA'' or ``AB/BA''-stacked bilayer graphene.
Intuitively, the point-group origin of the lattice coincides with the highest-symmetry point in the AA region. As we considered a small twist $\theta \ll 1$ starting from an AA-stacked bilayer, one might assume the point-group origin always coincides with the twisting center. This is not generally true. More specifically, of the four distinct cases of lattices we considered (type I vs.\ II angles $\theta \ll 1$ and twisting about a common hexagon center vs.\ a carbon site), it is true only for three of the four cases. The exceptions are type II structures generated by twisting about a common carbon site. As discussed above, the twisting center in this case has only $D_3$ symmetry, but the point group of the lattice has to be $D_6$ due to the existence of additional, aligned hexagon centers. This implies there must be a ``better'' AA region elsewhere centered at the aligned hexagon centers. We are thus forced to conclude that there are multiple AA regions in one primitive unit cell (Fig.\ \ref{fig:commensurate}). This is, in fact, nothing but a manifestation of the described approximate translation symmetries with pitch $L'<L$.
For an example, see Fig.\ref{fig:commensurate}(g).

\begin{table}[htp]
\caption{Possible commensurate structures for a given angle $\theta$ parameterized by coprime positive integers $m, r$. Structures with $\rm {gcd(r, 3)}=1$ are dubbed ``type I,'' and those with $\rm {gcd(r, 3)}=3$ are called ``type II.''
We let $L(\theta)$ be the {\it exact} \moire lattice constant, and $L'(\theta) \equiv a/ (2 \sin( \theta/2))$ be the effective lattice constant when $\theta \ll 1$.
Center refers to the twisting center, and PG the point group  of the lattice.
``Sym. rep.'' denotes the representations of $C_3$ (about the point-group origin) furnished by the Dirac points sitting at the \moire K points, and the symbol $\cup$ indicates how they are distributed across the two decoupled valleys when $\theta \ll 1$.
The case of ``conjugated'' structures, with twist angles $\pi/3 - \theta$, are discussed in Appendix \ref{sec:conj}.
\label{tab:summary}
}
\begin{center}
\begin{tabular}{ccccc}
\hline \hline
Type & $L(\theta)/L'(\theta)$ & Center & PG & Sym.\ rep.\ \\
\hline
\multirow{2}{*}{I}
& \multirow{2}{*}{$r$}& site & D$_3$ & $ (1,\omega) \cup (1,\omega^*)  $\\
~& ~  & hexagon & D$_6$ & $ (\omega,\omega^*) \cup (\omega,\omega^*) $\\
\hline
\multirow{2}{*}{II} & \multirow{2}{*}{$r/\sqrt{3}$}  & site &  D$_6$ & $ (\omega,\omega^*) \cup (\omega,\omega^*) $\\
~&  ~ & hexagon &  D$_6$ & $ (\omega,\omega^*) \cup (\omega,\omega^*) $\\
\hline \hline
\end{tabular}
\end{center}
\end{table}

\section{Symmetry representations of Dirac points in commensurate structures
\label{sec:comsym}}
Having discussed the geometrical aspects of commensurate TBGs, we now focus on the symmetry representations at the \moire K points. Understanding the symmetry representations serves as preliminaries for the construction of tight-binding models, which requires the identification of the correct real-space orbitals (i.e., Wannier functions).

As a warm up, let us first consider a monolayer graphene with the same lattice vectors $\vec a_{1,2}$ described before. In addition, we place the carbon atoms at $\frac{1}{2} (\vec a_1 + \vec a_2) \mp \frac{1}{6} (\vec a_1 - \vec a_2)$. Among the many spatial symmetries, the monolayer is, in particular, symmetric under a $ C_3^{\rm H}$ rotation about the origin (i.e., a hexagon center), as well as the lattice translations $T_{\vec a_{1,2}}$. Consider the Dirac-point Bloch states $|\psi_{\vec K}^\sigma\rangle$ at the momentum $\vec K = (\vec b_1 + \vec b_2)/3$, where $\vec b_{1,2}$ are the monolayer reciprocal lattice vectors. Here, $\sigma = \pm1$ denotes the sublattice degrees of freedom.
One can verify that
\beq \label{eq:C3Hrep}
C_3^{\rm H} |\psi_{\vec K}^\sigma\rangle = |\psi_{\vec K}^\sigma\rangle \omega^{\sigma},
\eeq
where we let $\omega = e^{i 2 \pi /3}$. Similarly, we also have $C_3^{\rm H} |\psi_{\vec K'}^\sigma\rangle = |\psi_{\vec K'}^\sigma\rangle \omega^{-\sigma}$.

Aside from the hexagon centers, the monolayer is also $C_3$-symmetric about the carbon sites. These rotations, denoted as $C_3^{\rm C}$, are simply the product of lattice translations and $C_3^{\rm H}$. For instance, check that $C^{\rm C}_3 \equiv T_{\vec a_1}T_{\vec a_2} C_3^{\rm H}$ leaves the site $(\vec a_1 + 2 \vec a_2)/3$ invariant.
As $T_{\vec a_1}T_{\vec a_2} |\psi_{\vec K}^\sigma\rangle = |\psi_{\vec K}^\sigma\rangle \omega^* $, we have:
\beq \label{eq:C3Crep}
C_3^{\rm C} |\psi_{\vec K}^\sigma\rangle =  T_{\vec a_1}T_{\vec a_2} C_3^{\rm H}  |\psi_{\vec K}^\sigma\rangle
=  |\psi_{\vec K}^\sigma\rangle  \omega^{\sigma-1}.
\eeq
The case for $\vec K'$ is fixed by TR symmetry, which commutes with rotations:
$C_3^{\rm C} |\psi_{\vec K'}^\sigma\rangle =  |\psi_{\vec K}^\sigma\rangle  \omega^{-(\sigma-1)}$.
Curiously, this simple relation between Eqs.\ \eqref{eq:C3Hrep} and  \eqref{eq:C3Crep} will be a recurrent motif in this section.

We are now ready to consider the TBG case. The analysis only requires two additional pieces of data: (i) the momentum mapping tabulated in Fig.\ \ref{fig:folding}, and (ii) that the Bloch states near charge neutrality are essentially dressed version of the monolayer Dirac points. More concretely, consider a type I structure, then
\begin{equation}\begin{split}\label{eq:KExpand}
| \psi_{\vec K_{\rm m}} ^{ \sigma}, I_z=+ \rangle \propto&   |\psi_{\vec K_{\theta}}^\sigma\rangle + \dots;\\
| \psi_{\vec K_{\rm m}} ^{ \sigma }, I_z = -\rangle \propto&   |\psi_{\vec K'}^\sigma\rangle + \dots,
\end{split}\end{equation}
where $I_z=\pm 1$ denotes the two effectively decoupled valleys. Note that only states with the same symmetry properties as the leading term can enter into the ellipsis.
The case for type II structures is essentially identical, but with $\vec K' \mapsto \vec K$ in $| \psi_{\vec K_{\rm m}} ^{\sigma}, - \rangle $.

To proceed with the analysis, however, we must first specify our twisting center, as the choice would determine the exact point-group symmetries of the lattice.
If we choose the twisting center to be an aligned hexagon center, we can reconcile the $C_3$ rotation about the AA region as $C_3^{\rm AA} = C_3^{\rm H}$ (here, interpreted as the direct sum of the single-particle symmetry matrix of the two layers)\footnote{
As we discussed in Sec.\ \ref{sec:review}, for certain commensurate lattices there can be multiple AA regions within each primitive unit cell. In such cases, we define $C_3{\rm AA}$ to be a three-fold rotation about the highest-symmetry point.
}. From Eqs.\ \eqref{eq:C3Hrep} and \eqref{eq:KExpand}, we have
\begin{equation}\begin{split}\label{eq:D6Rep}
C_3^{\rm AA} | \psi_{\vec K_{\rm m}} ^{ \sigma},+ \rangle =& | \psi_{\vec K_{\rm m}} ^{ \sigma},+ \rangle \omega^{\sigma}; \\
C_3^{\rm AA} | \psi_{\vec K_{\rm m}} ^{\sigma}, - \rangle =&  | \psi_{\vec K_{\rm m}} ^{\sigma}, - \rangle \omega^{-\sigma},
\end{split}\end{equation}
i.e., the representation of $C_3^{\rm AA}$ at $K_{\rm m}$ is $( \omega,\omega^* ) \cup ( \omega, \omega^*)$ for the four states near charge neutrality. Here, and in the following, we use $\cup$ to indicate how the representations are distributed across the two valleys.
The same representation content can be found for $K_{\rm m}'$, and the same conclusion holds for type II structures.

Alternatively, consider placing the twisting center at an aligned carbon site. As we have reviewed in Sec.\ \ref{sec:review}, in this setting we should treat types I and II structures separately. First, consider a type I lattice, for which $C_3^{\rm AA} = C_3^{\rm C}$.
The same analysis, but now combining Eqs.\ \eqref{eq:C3Crep} and \eqref{eq:KExpand}, leads to
\begin{equation}\begin{split}\label{eq:}
C_3^{\rm AA} | \psi_{\vec K_{\rm m}} ^{\sigma}, + \rangle =&| \psi_{\vec K_{\rm m}} ^{\sigma}, + \rangle \omega^{\sigma-1}; \\
C_3^{\rm AA}| \psi_{\vec K_{\rm m}} ^{\sigma}, - \rangle =& | \psi_{\vec K_{\rm m}} ^{\sigma}, - \rangle \omega^{-\sigma+1},
\end{split}\end{equation}
i.e., we now have the representations $( 1,\omega) \cup (1, \omega^*)$.
The case for conjugate type II structures is a bit more intriguing, and is covered in Appendix \ref{sec:conj}.

The conclusions from the analysis above can be summarized as follows (Table \ref{tab:summary}): if the point-group of the commensurate structure is $D_6$, the representation of $C_3^{\rm AA}$ at K$_{\rm m}$ is $( \omega,\omega^*) \cup (\omega,\omega^*)$; alternatively, if the point-group is $D_3$, then it becomes $( 1,\omega) \cup (1,\omega^*)$.
Though simple to derive, these results constrain the possible form of the tight-binding models. Indeed, as is pointed out in Refs.\ \onlinecite{NoahLiang, Mar2018, Oskar}, by taking the representations at the other high-symmetry momenta into account, one finds that the only possible tight-binding models for the four nearly flat bands near charge neutrality must have orbitals centered at the \moire honeycomb sites (corresponding to the AB/BA regions).
However, the symmetry characters of the orbitals are {\it necessarily} different for $D_3$ vs.\ $D_6$, for otherwise one cannot reproduce the representations at $K_{\rm m}$. Indeed, for the $D_6$ case it was found in Ref.\ \cite{Mar2018} that the Wannier orbitals transform trivially under a three-fold rotation about their charge centers, whereas for the $D_3$ case the orbitals should transform under the $( \omega, \omega^*)$ representation \cite{NoahLiang, Oskar}.

Taken at face value, these results are individually self-consistent and there is no contradiction between them, but they appear different. How do we understand such discrepancy? There are two possibilities: (i)
The $D_3$ and $D_6$ structures are physically distinct, and their different symmetry properties lead to distinct electronic behavior; (ii) The two classes of commensurate structures are ultimately described by the same effective theory, and therefore the apparent distinction in symmetry representations is simply an artifact of the commensurate approximants and has little physical implication.

Curiously, for TBG both possibilities are applicable, but they are operative in different parameter regimes. As is pointed out in Ref.\ \onlinecite{Mele2010}, for large twist angles case (i) applies, and the system is generically gapped or gapless at charge neutrality depending on the symmetry setting; for small twist angles, however, multiple studies have pointed out that the key electronic properties of the system at the meV energy scale are universal and become independent of the {\it exact} geometric details of the system \cite{Neto2007,Bistritzer2011, Castro-Neto2012,Jung2014}, i.e., case (ii) applies. In particular, this implies the distinction between $D_3$ and $D_6$ structures cannot matter in a proper treatment of the electronic properties of the system, unless one is interested in quantities resolved to the $\mu$eV scale. In other words, the symmetries of the effective theory must contain at least those of {\it both} $D_3$ and $D_6$. But since $D_3$ is a strict subgroup of $D_6$, it suffices to consider $D_6$ commensurate lattices if one is interested in a more microscopic treatment of the problem.

We remark that, if one desires, one can also instead study $D_3$ commensurate, small-angle TBG structures \cite{
NoahLiang, Oskar, KoshinoLiang}. However, in such a setup, $C_6$, an exact symmetry in the $D_6$ case and a good symmetry of the effective theory, would become an {\it approximate} symmetry.
If one incorporates this approximate symmetry in the analysis, the problem will be enhanced to the $D_6$ case; alternatively, if $C_6$ is completely ignored, the Dirac points at charge neutrality would lose symmetry protection (even in the limit of exact valley charge conservation), i.e., in such strictly $D_3$ treatments the Dirac points are not robust features of the models, but rather appear as accidental energetic features which require fine-tuning of model parameters, say, by forcing certain symmetry-allowed terms to vanish \cite{KoshinoLiang}.


%
%

\section{On the continuum theory \label{sec:cont}}
In Secs.\ \ref{sec:review} and \ref{sec:comsym}, we have reviewed the known classes of commensurate TBG structures, as well as how the symmetries are represented by the electronic states on such lattices.
A key lesson learnt in the past decade of studies on TBG is that, in the limit of small twist angle, the precise form of the lattice realization becomes irrelevant and the system is well-described by a continuum theory with two decoupled, TR-related sectors corresponding to the two microscopic valleys. Specifically, the Hamiltonian for a single valley takes the form $\hat H_{\rm Cont.} = \hat H_{\rm Dirac} + \hat H_{\rm T}$, where $\hat H_{\rm Dirac}$ encodes the Dirac dispersion originating from the monolayer K points, and the coupling between the two layers are given by
\begin{equation}\begin{split}\label{eq:CTh}
\hat H_{\rm T} =&  \int_0^{\Lambda} d^{2} \vec k \, \hat  \psi_{+{\rm b};  \vec k}^\dagger \,  T_{\vec q_1}\,  \hat  \psi_{+{\rm t};  \vec k + \vec q_1}\, + {\rm h.c.}\\
&~~~+ \text{symmetry related terms},
\end{split}\end{equation}
where ${\rm t}$ and ${\rm b}$ respectively denote the top and bottom layers, and $\Lambda$ is a high-momentum cutoff.
The momentum $\vec q_1 \equiv R_{-\theta/2} \vec K- R_{\theta/2}\vec K$ characterizes the momentum transfer between the electronic degrees of freedom of the two layers  \cite{Bistritzer2011}.
Here, we rotate the top layer by $\theta/2$ and the bottom by $-\theta/2$ to construct a system with a total twist angle of $\theta$. The resulting \moire pattern with this setup is shown in Fig.\ \ref{fig:commensurate}.

In Ref.\ \onlinecite{Bistritzer2011}, the coupling matrix is given by $T_{\vec q_1} = w (\sigma_0 + \sigma_1)$, but more generally one can take  $T_{\vec q_1} = w_0 \sigma_0 + w_1 \sigma_1$ without breaking any symmetries \cite{Mele2010, Mar2018, You2018, KoshinoLiang}. In fact, as in any effective theory, one can free oneself from the microscopic problem and instead allow for the presence of any physical terms, unless they are symmetry-forbidden\footnote{
If the symmetry is emergent/ approximate, as for the spatial symmetries in the present context, symmetry-breaking terms are in principle allowed, but their associated energy scale is suppressed below the one of interest.
}
or involve high-order processes and are therefore energetically suppressed.
Therefore, it is important to understand the symmetries of the theory. As we have alluded to, aside from TR and valley charge conservation, the continuum theory will have the spatial symmetries of a $D_6$ commensurate realization, described by the wallpaper group 17.
The concrete representations for valley-preserving symmetries, as listed in Ref.\ \onlinecite{Mar2018}, are reproduced below:
\begin{equation}\begin{split}\label{eq:SymRep}
(\hat C_6 \hat {\mathcal T}) \hat \psi_{I_z,  \mu}^\dagger (\vec k) (\hat C_6 \hat {\mathcal T})^{-1} =\,& \hat \psi_{I_z,  \mu}^\dagger (- C_6\vec k) \left( e^{-i \frac{2 \pi }{3}\sigma_3 I_z } \sigma_1 \right) ;\\
\hat M_y \hat \psi^\dagger_{I_z,  \mu} (\vec k) \hat M_y^{-1} =\,& \hat \psi^\dagger_{I_z,  M_y[\mu]} (M_y \vec k)  \sigma_1 .
 \end{split}\end{equation}
where $\mu = {\rm t}, {\rm b}$, and $I_z = \pm$ denotes the two valleys. Note that $M_y$, which as mentioned is in fact a two-fold rotation in 3D, is the only symmetry which flips the two layers, i.e., $M_y [{\rm t}] = {\rm b}$ and vice versa. In particular, note that while $C_6$ and TR $\mathcal T$ individually flip the valleys, $\hat C_6 \hat I_z \hat C_6^{-1} = \hat {\mathcal T} \hat I_z {\mathcal T}^{-1} =  - \hat I_z$, their product is a symmetry of the single-valley problem.
One can check explicitly that $\hat H_{\rm Cont.} $ is invariant under these symmetries. In particular, this implies $\hat H_{\rm Cont.} $  is $C_6$ symmetric when the other valley is taken into account.

As $(\hat C_6 \mathcal T)^2 = \hat C_3$, the representation of $\hat C_3$ is also fixed by Eq.\ \eqref{eq:SymRep}. Let $\mathcal U(C_6 \mathcal T) = e^{-i \frac{2 \pi }{3}\sigma_3 I_z } \sigma_1$, then
\begin{equation}\begin{split}\label{eq:C3C6T}
\mathcal U(C_3) =\mathcal U(C_6 \mathcal T) \mathcal U^*(C_6 \mathcal T) = e^{i \frac{2\pi}{3} \sigma_3 I_z}.
\end{split}\end{equation}
Therefore, at the \moire K point one finds that the symmetry eigenvalues of $C_3$ are $( \omega, \omega^*) \cup ( \omega,\omega^*)$ for the four states near charge neutrality, which corresponds to the $D_6$ case analyzed in Sec.\ \ref{sec:comsym}
\footnote{
For simplicity, we drop the superscript ``AA'' here, with the rotation centers understood to be taken about the point-group origin. Also, we have implicitly used the fact that the states near charge neutrality descends from the microscopic Dirac points; for states away from neutrality, other representations become possible, as the spatial symmetry also permutes the momenta which are not exactly at the microscopic K point.
}.

As the $D_6$ and $D_3$ commensurate structures have the same effective theory, one may wonder how to recover the other set of symmetry representations derived from the $D_3$ lattices \cite{NoahLiang, Oskar, KoshinoLiang}. To this end, let us first restrict our attention to the spatial symmetries that are {\it exact} on a $D_3$ lattice, $C_3$, $M_x \equiv M_y C_6^3$.
Recall $\mathcal T$ and $M_x$ anticommutes with $I_z$ whereas $C_3$ commutes, and
note the following group relations involving $C_3$:
\begin{equation}\begin{split}\label{eq:C3Rel}
\begin{array}{rlrl}
 C_3  I_z  C_3^{-1} &=  I_z; &
 {\mathcal T}  C_3  {\mathcal T}^{-1} &= C_3;\\
 M_x  C_3  M_x^{-1} &=   C_3^2;&
 C_3^3 &=  1.
\end{array}
\end{split}\end{equation}
Let $\tilde C_3 \equiv e^{i \frac{2\pi}{3} I_z} C_3$. One can check that Eq.\ \eqref{eq:C3Rel} is equally satisfied by $C_3 \mapsto \tilde C_3$. However, in a single valley, say $I_z = +1$, we have ${\rm eig}(\mathcal U(\tilde C_3)) = \omega \times {\rm eig}(\mathcal U(C_3)) $. This toggles between the two set of $C_3$ representations found in Sec.\ \ref{sec:comsym} for $D_3$ and $D_6$ cases.

Physically, the redefinition of $C_3 \rightarrow \tilde C_3$ amounts to a redefinition of the rotation center at the microscopic lattice scale. To see why, we simply note that the emergent ${\rm U}_{\rm v}(1)$ symmetry can be reconciled with the microscopic lattice translation $T_{\vec a}$, which becomes an effective ${\rm U}(1)$ when acting on the slowly varying degrees of freedom in the effective theory. Therefore, attaching the valley-dependent phase $ e^{i \frac{2\pi}{3} I_z} $ to $C_3$ can be physically interpreted as multiplying by a small translation $T_{\vec a}$.

If $C_6$ symmetry is truly absent in the theory (i.e., not even an approximate symmetry), the choice between $C_3$ and $\tilde C_3$ above is completely arbitrary, and, so long as ${\rm U}_{\rm v}(1)$ is a good symmetry, physical observables should not depend on this choice.
For instance, in the discussions on Wannier functions in Ref.\ \onlinecite{NoahLiang, Oskar, KoshinoLiang}, the orbital character of $(\omega, \omega^*)$ under $C_3$ rotation can equally be changed into the trivial one as long as one admits that $U_{\rm v}(1)$ is a good emergent symmetry (e.g., on the tight-binding model).
That said, the commensurate calculations in Ref.\ \onlinecite{NoahLiang, Oskar} are performed on microscopically well-defined tight-binding models, and because of that one loses the exact ${\rm U}_{\rm v}(1)$ symmetry.
In such calculation, one has to make a choice between the two set of possible $C_3$ representations discussed in Sec.\ \ref{sec:comsym}, depending on whether the point-group origin is placed at a common carbon site ($D_3$) or a common hexagon center ($D_6$).
The choice of lattice realization and the reported symmetry representations are therefore internally consistent in Refs.\ \onlinecite{NoahLiang, Oskar}.
In contrast, in the calculation in Ref.\ \onlinecite{KoshinoLiang}, which starts from the effective theory instead of a commensurate calculation, the choice of $C_3$ vs.\ $\tilde C_3$ representation is arbitrary.

Importantly, in the preceding paragraph on $C_3$ vs.\ $\tilde C_3$, as well as in Refs.\ \onlinecite{NoahLiang, Oskar, KoshinoLiang}, the valley-preserving symmetry $C_6\mathcal T$ is ignored.
$C_6\mathcal T$ is an exact symmetry of the continuum theory, and is also an excellent approximate (if not exact) symmetry of any small-angle commensurate realizations.
No matter which microscopic regularization one prefers, it is desirable to identify its representation in the theory (even if not an {\it exact} symmetry).
As we argue below, this requirement picks out a preferred choice for $C_3$ vs.\ $\tilde C_3$: recall, as symmetries, we have $(C_6\mathcal T)^2 = C_3$. Naturally, we demand their single-valley representations to follow $\mathcal U_+(C_3) = \mathcal U_+(C_6 \mathcal T) \mathcal U_+^*(C_6 \mathcal T) $, where $+$ denotes the $I_z=+$ valley. As ${\rm det} \left(  \mathcal U_+(C_6 \mathcal T) \mathcal U_+^*(C_6 \mathcal T) \right) = 1$, this forces the two eigenvalues of $\mathcal U_+ (C_3)$ to form a conjugate pair, i.e., one should arrive at ${\rm eig}(\mathcal U_+ (C_3)) = ( \omega, \omega^*)$. Consequentially, we conclude that, in a microscopic calculation, one should place the point-group origin at an aligned hexagon center in order to define the representations for all the relevant spatial symmetries of the system. For a $D_6$ lattice, this choice is automatic, and all the mentioned spatial symmetries are exact; for a $D_3$ lattice, however, this choice does not coincide with the point-group origin defined using {\it exact} symmetries, but the system would still be approximately invariant under the $D_6$ symmetries about the aligned hexagon center.

To summarize, the continuum theory, well-known to capture all the salient features of the electronic band structures of small-angle TBG, has a $C_6$ rotation symmetry. To incorporate this symmetry one naturally arrives at the set of symmetry representations realized as in a microscopic lattice with $D_6$ point-group symmetries. This suggests that the most natural commensurate lattices to study would be type I structures with the twisting center chosen to be an aligned hexagon center, which has all the exact spatial symmetries (point group and translation) of the continuum theory. If a different class of microscopic lattice is picked, say those chosen in Refs.\ \onlinecite{NoahLiang, Oskar}, then the excellent emergent symmetry of $C_6$ is masked, and it becomes difficult to justify, for instance, the robustness of the Dirac points at charge neutrality. We remark that this problem is particularly severe in the presence of perturbations which break the layer-exchanging symmetries (say, $M_x$), say when one applies a perpendicular electric field: while it is known that the Dirac points remain stable against such perturbation \cite{Neto2007}, in the microscopic $D_3$ set up there is no {\it exact} symmetry reason to expect any band degeneracy at charge neutrality.

%

\section{Remarks on magic angles} \label{sec: remark_magic_angle}

In this section we make further remarks on the notion of the magic angle, which in the theoretical literature is defined as the angle where the Dirac speeds at $K$ and $K'$ vanish.

Generically, symmetries constrain the single-valley band structure, say, near the $K$ point, to be described by \cite{Cao2018a, Mar2018}
\beq \begin{split}
H(\vec k)=&
v_D
\left(
\begin{array}{cc}
0 & k_x-ik_y\\
k_x+ik_y & 0
\end{array}
\right)\\
& ~~+
\frac{1}{2m}
\left(
\begin{array}{cc}
0 & (k_x+ik_y)^2\\
(k_x-ik_y)^2 & 0
\end{array}
\right)
+\mc{O}(k^3)
\end{split}
\eeq
Close to the magic angle, $mv_D$ is smaller than the size of mBZ, so the above Hamiltonian gives one Dirac point right at the $K$ point, as well as three satellite Dirac points away from the $K$ point. Right at the magic angle, these satellite Dirac nodes all come to $K$ and $K'$, so that these two points show chiral quadratic band touching. This is similar to the phenomenon of trigonal warping as in AB-stacked bilayer graphene.

One may wonder where these satellite Dirac points come from, because they are absent from the band structure calculations at larger twist angles. There are two possible scenarios. First, it is possible that new Dirac nodes emerge from other points (such as the $\Gamma$ point) in mBZ as the magic angle is approached, while the nearly flat bands remain isolated from other bands by a gap. Upon further approaching the magic angle, some of these new Dirac nodes move towards the $K$ and $K'$ points and become the satellite Dirac nodes. In this scenario, the net chirality of the nearly flat bands does not change upon approaching the magic angle. The other possibility is that the energy gap between these nearly flat bands and other bands closes and reopens when the magic angle is approached. When this happens, some new Dirac nodes can appear in mBZ and move to $K$ and $K'$ as the magic angle is approached. In this scenario, the net chirality of the nearly flat bands may change upon approaching the magic angle.

Depending on the parameters in the model, either scenario can be realized in a band structure calculation based on the continuum model. However, experimentally there appears to be an energy gap that is much larger than the resulting gap from the continuum model \cite{PabloPRL, Cao2018a, Cao2018}, which can be a consequence of interaction effects and/or lattice relaxation that are ignored. This also indicates that in the real experiments the gap between the nearly flat bands and other bands never closes, so if the magic angle defined above can be arrived, the first of the above scenarios should be realized. However, we would like to point out that operationally the magic angle may be defined in other ways, which is, a priori, unrelated to whether the Dirac speeds at $K$ and $K'$ vanish. For example, the magic angle can be defined to be where the ratio to the band gap and the band width of the nearly flat bands is maximized, or simply to be where the gap is the largest.

\section{More on Wannier obstruction} \label{sec: relation_obstruction}

\subsection{Mirror and chirality}

In the introduction we have pointed out the existence of two obstructions to constructing exponentially localized Wannier functions for the two nearly flat bands of a single valley and spin in TBG \cite{Mar2018}:
\begin{itemize}

\item[]1. Mirror-eigenvalue obstruction: The mirror $M_y$ eigenvalues at $M$ (or $\Gamma$) is $\pm 1$.

\item[]2. Chirality obstruction: The entire Brillouin zone for a single valley has a nonzero net chirality. More precisely, the two Dirac points of the single-valley band structure have the same chirality.

\end{itemize}

In this section, we will elaborate on the relation between these two Wannier obstructions. It turns out that these two obstructions are equivalent in the context of TBG, as long as the system preserves the $M_y$ symmetry. When these obstructions are initially realized on a setting that has the $M_y$ symmetry, the chirality obstruction remains even if the $M_y$ symmetry is broken later, because the chirality is a discrete object that should not change upon breaking the mirror symmetry (at least weakly).

This observation is significant not only conceptually, but also practically. This is because there is always an intrinsic phase ambiguity associated with the Bloch wave function of a band structure, and it requires a smooth choice of the Bloch wave functions across the mBZ to determine the chirality. However, it takes some efforts to obtain such a smooth basis of Bloch wave functions. The above observation then greatly simplifies the problem of checking the chirality obstruction in TBG, since now one only has to check the mirror eigenvalues at high symmetry points in an $M_y$ symmetric setting, which does not require choosing a smooth basis across the entire mBZ.

Below we only sketch the logic to show the above statement, and leave the details in Appendix \ref{app: relation_obstruction}. The chirality is only contributed by the gapless points in mBZ, so we can focus on the an {\em open} region of the mBZ that covers the gapless points. Unless very close to the magic angle, the only gapless points are the $K$ and $K'$ points. Upon approaching the magic angle, assuming the first scenario of generating satellite Dirac nodes discussed in Sec. \ref{sec: remark_magic_angle} is realized, the net chirality will not change compared to the case before these satellite Dirac nodes appear, so we can always obtain the net chirality by looking at the Dirac nodes at $K$ and $K'$.

In appendix \ref{app: relation_obstruction}, we will show that there exists a smooth basis of Bloch wave functions so that the action of $C_2\mc{T}$ is
\beq
\psi(\vec k)\rightarrow\sigma_x\mc{K}\psi(\vec k)
\eeq
where $\psi(\vec k)$ is a two-component operator that annihilates an electron at momentum $\vec k$ in the two nearly flat bands, and $\mc{K}$ stands for complex conjugation. In this basis, the first-quantized Hamiltonian can be written as
\beq
H(k)=n_0(\vec k)+n_1(\vec k)\sigma_x+n_2(\vec k)\sigma_y
\label{eqn:H}
\eeq
The chirality is given by the winding of $(n_1(\vec k), n_2(\vec k))^T$.

Furthermore, it is shown that in this basis the action of $M_y$ can be chosen as
\beq
\psi(\vec k)\rightarrow\sigma_x\psi(\vec k)
\eeq
if the mirror eigenvalues at $M$ are opposite, and as
\beq
\psi(\vec k)\rightarrow\eta_M\psi(\vec k)
\eeq
if the mirror eigenvalues at $M$ are both $\eta_M$. For two momenta related by $M_y$, say, $\vec k$ and $\vec k'$, $M_y$ requires
\beq \label{eq: mirror-constraint-1}
n_1(\vec k)=n_1(\vec k'),
\quad
n_2(\vec k)=-n_2(\vec k')
\eeq
if the mirror eigenvalues at $M$ are opposite, and
\beq \label{eq: mirror-constraint-2}
n_1(\vec k)=n_1(\vec k'),
\quad
n_2(\vec k)=n_2(\vec k')
\eeq
if the mirror eigenvalues at $M$ are identical.

To check the net chirality, now one can consider a small closed loop around each of $K$ and $K'$. It is straightforward to see that the windings around these two loops are the same if (\ref{eq: mirror-constraint-1}) holds, and they are opposite if (\ref{eq: mirror-constraint-2}) holds. This implies that having opposite (identical) mirror eigenvalues at $M$ is equivalent to having nonzero (zero) net chirality in mBZ.

The above claim stating that the mirror-eigenvalue obstruction and chirality obstruction are equivalent applies to any two-band system that has an odd number of mirror-related pairs of Dirac nodes. If there are an even number of mirror-related pairs of Dirac nodes, a nonzero net chirality still implies the mirror-eigenvalue obstruction.  However, the converse is not true: the existence of the mirror-eigenvalue obstruction does not immediately imply a nonzero net chirality. To settle down the net chirality in this case, one can divide the entire Brillouin zone into two halves that are mirror-related, and then check the total chirality of one of the two halves.

\subsection{An Alternate Viewpoint: the Flipped Haldane model}

Let us now provide a more physical argument for the obstruction to realizing a tight-binding model for a single valley of the nearly flat band. Crucial to this argument is that we do not augment the model with additional bands. Let us begin by assuming the opposite---that there is a tight-binding model with a single orbital on the honeycomb lattice that indeed realizes the valley filtered band structure as given by (\ref{eqn:H}) (where the two component wavefunctions now refer to fixed sublattices in a tight-binding model), including the two Dirac points at the mini BZ K$_{\rm m}$, K$_{\rm m}$' points, with the same chirality. Recall, the conventional situation, as realized in graphene, is to have Dirac points with opposite chiralities. With this setup consider adding a staggered potential $\pm m$, opposite on the two honeycomb sublattices. This will induce a gap at the Dirac points, since $C_2$ symmetry is broken. In particular, the term has no momentum dependence since it comes from an {\em onsite}  term. In addition, if both Dirac points have the same chirality, then the contributions to the Chern number of the disconnected bands will add to $C=\pm1$ (the opposite valley has $C=\mp 1$). Note, if we had the conventional case of opposite chiralities, the contributions cancel as in graphene with a staggered onsite potential. One can view this as a ``flipped'' Haldane model, where the staggered potential produces the Chern band and the the trivial inversion breaking insulator is obtained by the second neighbor Haldane hopping term \cite{Haldane}.

This produces the following contradiction - at strong onsite potential on the honeycomb lattice, an atomic insulator is obtained, which is not compatible with a finite Chern number band, and which cannot be reduced to localized Wannier functions.

Note, if we allowed for additional sites in the tight-binding model, the staggered potential is no longer a purely onsite term, and the previous contradiction does not hold. Obtaining such an augmented model which resolves the anomaly is an important future goal.

 \section{Resolving the obstruction} \label{sec: resolve}

We have argued that, in the presence of $U_v(1)$ valley charge conservation and the six-fold rotation $C_6$ (which combines with TR to generate the $C_2\mathcal T$  symmetry required in defining chirality), there is an obstruction to construct symmetric, well-localized Wannier functions for the four nearly flat bands near charge neutrality (spin degeneracy ignored in the counting). Such Wannier obstructions are reminiscent of that in topological insulators \cite{Z2Wannier}, and are an integral part in the understanding of the physical properties of the system.

Similar to the parallel discussions for topological insulators, to construct well-localized Wannier functions one has to forgo some of the symmetries in the problem.
In Ref.\ \onlinecite{Mar2018}, we describe a construction which first forgoes $U_v(1)$ but retains $C_6$; in contrast, in Refs.\ \onlinecite{NoahLiang, Oskar, KoshinoLiang}, $C_6$ is completely ignored and $U_v(1)$ is kept\footnote{More accurately, there exists an identification of the valley-charge operator $I_z$ such that one can demand $U_v(1)$ symmetry in the resulting tight-binding model.}. At this level, both approaches may seem imperfect, in that the Dirac points at charge neutrality are not symmetry-protected robust features. However, in Ref.\ \onlinecite{Mar2018} we make one extra step and identify the concrete representation of the missing symmetry, generated by the valley charge operator $I_z$. This allows us to consistently restore $U_v(1)$ symmetry within the tight-binding Hilbert space, such that {\it all} symmetries become manifest in the final model.

Interestingly, the Wannier obstruction we identify here is tied to the $C_2 \mathcal T$ symmetry, which is non-local. Unlike the corresponding discussion for topological band structures protected by internal symmetries, say topological insulators \cite{Z2Wannier}, spatial symmetries are known to lead to new types of Wannier obstructions \cite{Fragile} which are less stable, but at the same time more intricate, than the conventional topological indices like Chern numbers. The obstructions we identify here are connected to the notion of ``fragile topology'' \cite{Fragile}. This can provide a new, unconventional avenue for resolving the obstruction, wherein the obstruction can be resolved by the addition of {\it trivial} bands.
We will leave the full exploration of this connection to a future work.

\subsection{A four-band model for the nearly flat bands}

Let us now discuss one simple way of resolving the obstructions. We can write down a tight-binding model that produces two  sets of bands, each of which has the topology characteristic of the nearly flat bands of a single valley of TBG. These bands will be split so energetics will single out the lower band which has the desired character, and preserves all symmetries of the continuum theory, including valley conservation symmetry. This is loosely analogous to tight-binding models of Chern insulators that produce a pair of opposite Chern bands, although individual bands cannot be captured within in a tight-binding model.

\begin{figure}[h]
\centering
\includegraphics[width=0.45 \textwidth]{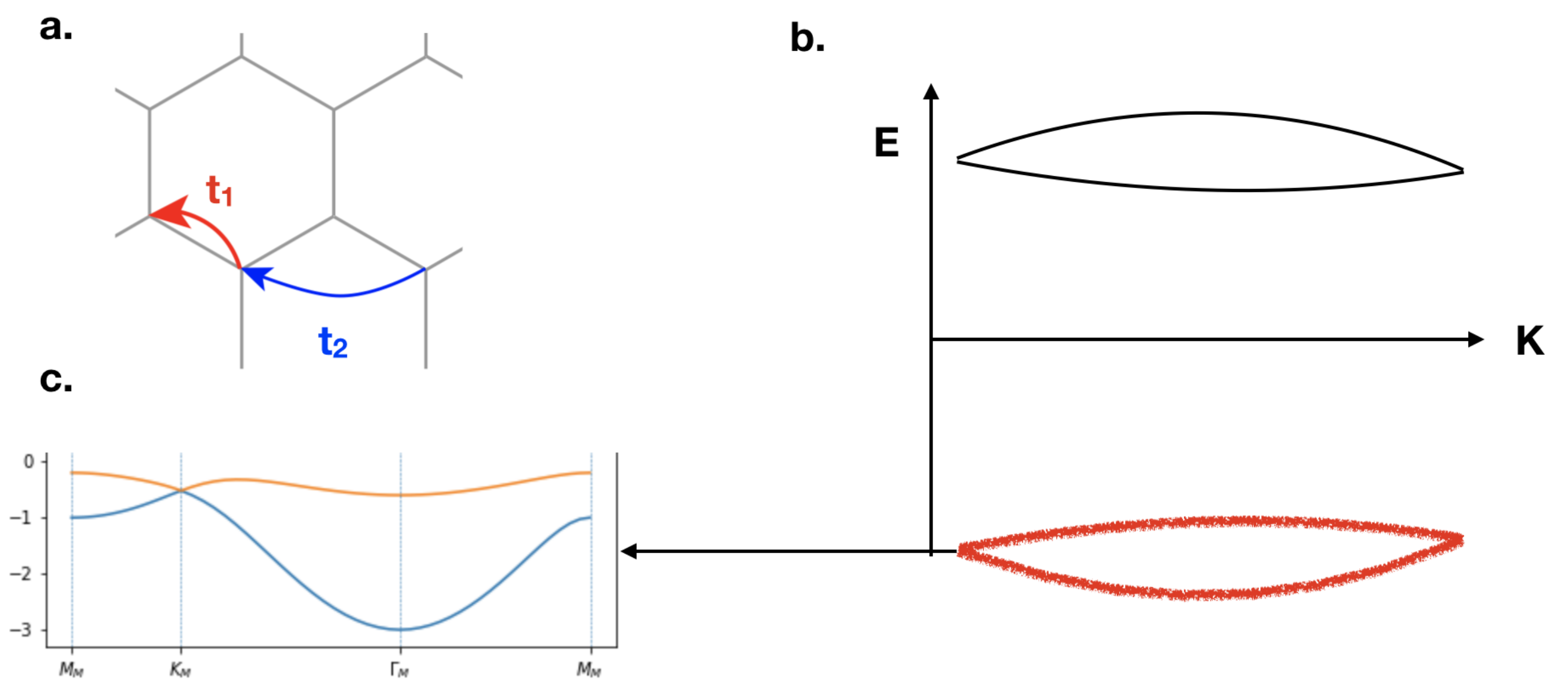}
\caption{(a)Tight-binding model of valley-resolved isolated Moire bands from a two-orbital model on the honeycomb lattice. (b) Schematic band structure of this model. Two connected sets of bands separated by a  bandgap result. The lower one, for example, captures the band structure of the nearly flat bands of twisted bilayer graphene, in a single valley while preserving all symmetry ($D_6$ point group). (c) Upon adding further symmetry-allowed perturbations, one can attempt to reproduce the band structure of twisted bilayer graphene. Dispersion shown is for parameters ${\hat{t}}_1 = 0.4\mu_0+0.6\mu_z, \, {\hat{t}}_2 = 0.1 i\mu_x$.}
\label{Fig:model}
\end{figure}

An example of such a tight-binding model that retains all symmetries is given below. Of course the price paid to the obstruction is an increase in the total number of bands. More specifically, we present a model with  four bands per valley, where the two lower energy bands are the relevant one.  Let us write the Hamiltonian in two parts: $H(k)=H_+(k)+H_-(k)$, where $H_-(k)=H_+(-k)$ will give us the two valleys with $\pm$ denotes the valley. To begin with, focus on the model for one valley $H_+(k)$, as shown in the Figure \ref{Fig:model}. The Hilbert space consists of two orbitals per site of the honeycomb lattice that transform trivially under  $C_3$ rotations:
\beq
c_{i}(\vec r)\rightarrow c_i(C_3\vec r)
\eeq
with $i=1, 2$ denoting the two orbitals. Under $C_2\mc{T}$ they transform as
\beq
c_{i}(\vec r)\rightarrow \mc{K}c_i(-\vec r)
\eeq
And under $M_y$ they transform as
\beq
c_{1}(\vec r)\rightarrow c_1(M_y\vec r),
\quad
c_2(\vec r)\rightarrow -c_2(M_y\vec r)
\eeq

Denoting Pauli operators in this onsite orbital space as $\mu_a$ with $a=0, 1, 2, 3$, where $\mu_0$ is the identity operator in this space, we consider a simple Hamiltonian involves nearest and next nearest neighbor couplings: ${\hat{t}}_1 = 0.4\mu_0+0.6\mu_z, \, {\hat{t}}_2 = 0.1 i\mu_x$. This four-band model (two orbitals per site and two sites in the unit cell) leads to the (schematic) band structure shown in Figure \ref{Fig:model}, where nearly flat bands are split by an energy of order unity. The lower bands, for example, has the required topology of Dirac points with identical chirality, that we wish to capture for TBG.  Also, the mirror-eigenvalues at $\Gamma$ are opposite for these two bands. Given that this model preserves all symmetries, one can perturb it with any symmetry allowed term, and the topological properties of the individual bands will be preserved as long as the gap between the two sets of bands remains open. Potentially, one can tune parameters to obtain agreement of the band structure with other calculations, and further to obtain relatively narrow bands and a large energy gap between the pairs of bands, which we leave to future work. The main point we emphasize here is that we have accurately captured the universal topological properties.

Now that we have a tight-binding model that captures all salient features of the band structure of TBG, we can in principle systematically incorporate interactions into the model. This is analogous to using a model of a Chern insulator to study the lowest Landau level problem, where the role the model of a Chern insulator is played by this model, and the role the lowest Landau level is played by the nearly flat bands of TBG. Ideally we would like to achieve narrow bandwidths and a large band gap between pairs of bands, that can be achieved by tuning the microscopic parameters, which we leave to future work.

Before closing, let us note the following points. First, this Hamiltonian is essentially the $I_z$ projector discussed in Ref. \cite{Mar2018}. Second, we would like to stress that all symmetries are preserved in this model, so that one can systematically incorporate interactions into the model while maintaining all universal topological aspects of the single-valley nearly-flat bands of TBG. This is in sharp contrast to Refs. \onlinecite{NoahLiang, Oskar, KoshinoLiang}, where  salient features of the band structures, such as the existence of Dirac nodes, require fine-tuning the Hamiltonian. Lastly, although the four bands taken together suffer no obstruction because they result from a well defined tight-binding model, both the two upper bands and the two lower bands individually suffer from the obstructions discussed earlier. In this sense, this model does not resolve the obstructions in the sense of fragile topology discussed in Ref. \onlinecite{Fragile}. We leave such a resolution for future work.

\section{Discussion} \label{sec: discussion}

In this paper, we first collected and reviewed  aspects of the band structure of TBG, based on which we clarified some confusing issues about different types of commensurate structures of TBG and their symmetries. In particular, there are two different types of commensurate structures according to the twist angle, and the {\em exact} point-group symmetry for these commensurate structures can either be $D_6$ or $D_3$. However, we emphasize that in the context of small-twist-angle TBG, the various excellent approximate symmetries that are responsible for the salient features of the band structure are more important than the exact symmetries, regardless whether the bilayer is commensurate or incommensurate, unless one is interested in ultralow energy scales that are inaccessible to the experiments to date.

In particular, we analyze the symmetry representations of the Dirac points in different commensurate structures of TBG, and we demonstrate how a $D_6$ point-group symmetry can protect the Dirac points in a single valley, while a $D_3$ point-group symmetry cannot. This is essentially the reason that forces us to consider a formulation of the problem that has manifest $D_6$ point-group symmetry (among with other symmetries, including $U_{\rm v}(1)$ and $\mc{T}$), since both theoretically and experimentally gapless Dirac points are observed in small-twist-angle TBG.

A powerful formulation that incorporates all these symmetries is the continuum model. We review this model and its symmetry properties. Along the way, we point out that an apparent disagreement on the $C_3$ representations of the Dirac points in our eariler paper Ref. \cite{Mar2018} and that in Refs. \cite{NoahLiang, Oskar, KoshinoLiang} is actually artificial: the difference is purely due to a different choice of the point-group center, and these results are consistent after converting to the same convention.

Next we make some remarks on the notion of the magic angle, which in the theoretical literature is commonly defined as the angle at which the Dirac speeds vanish. We discuss the two scenarios in which this can be realized, and their difference is whether the nearly flat bands of TBG touch other bands upon approaching the magic angle. Based on the experimental results, we point out that the nearly flat bands are likely to remain isolated from other bands when the magic angle is approached.

One important result from our approach is the existence of two obstructions to constructing well-localized Wannier functions with natural symmetry representations. One of the obstruction is related to the mirror eigenvalues at high symmetry points, and the other is related to the net chirality of the entire mBZ. We discuss the relation between these two obstructions in the context of TBG, and we point out that they are actually equivalent. Because the mirror-eigenvalue obstruction is much easiler to check than the chirality obstruction, this observation provides a neat simplifying tool to check the latter by examining the former.

Finally, at the price of doubling the number of bands, we present a tight-binding model that can reproduce all essential features of the band structure of the nearly flat bands of TBG. This model has two sets of bands that are well separated from each other, and each set presents all important topological aspects of the nearly flat bands of TBG. In particular, they preserve all symmetries. Starting from this model, one can systematically incorporate symmetry-allowed interactions to the model. In this model, both sets of the bands suffer from the Wannier obstructions, so one can view this model as one that removes the obstruction of the nearly flat bands of TBG by adding it another copy that carries the opposite obstruction. It is more desirable to develop a model that resolves the obstruction by adding some bands that carry no obstruction at all, which can show these obstructions are fragile obstructions. We leave this for future work.

\begin{acknowledgements}

We thank Shiang Fang, Liang Fu, Oskar Vafek, Cenke Xu and  Noah Yuan for discussions and/or helpful email correspondence. TS is supported by a US Department of Energy grant DE-SC0008739, and in part by a Simons Investigator award from the Simons Foundation. AV was supported by a Simons Investigator award and by NSF-DMR 1411343.
\end{acknowledgements}

\bibliography{TwBLGMott}

\begin{thebibliography}{48}%
\makeatletter
\providecommand \@ifxundefined [1]{%
 \@ifx{#1\undefined}
}%
\providecommand \@ifnum [1]{%
 \ifnum #1\expandafter \@firstoftwo
 \else \expandafter \@secondoftwo
 \fi
}%
\providecommand \@ifx [1]{%
 \ifx #1\expandafter \@firstoftwo
 \else \expandafter \@secondoftwo
 \fi
}%
\providecommand \natexlab [1]{#1}%
\providecommand \enquote  [1]{``#1''}%
\providecommand \bibnamefont  [1]{#1}%
\providecommand \bibfnamefont [1]{#1}%
\providecommand \citenamefont [1]{#1}%
\providecommand \href@noop [0]{\@secondoftwo}%
\providecommand \href [0]{\begingroup \@sanitize@url \@href}%
\providecommand \@href[1]{\@@startlink{#1}\@@href}%
\providecommand \@@href[1]{\endgroup#1\@@endlink}%
\providecommand \@sanitize@url [0]{\catcode `\\12\catcode `\$12\catcode
  `\&12\catcode `\#12\catcode `\^12\catcode `\_12\catcode `\%12\relax}%
\providecommand \@@startlink[1]{}%
\providecommand \@@endlink[0]{}%
\providecommand \url  [0]{\begingroup\@sanitize@url \@url }%
\providecommand \@url [1]{\endgroup\@href {#1}{\urlprefix }}%
\providecommand \urlprefix  [0]{URL }%
\providecommand \Eprint [0]{\href }%
\providecommand \doibase [0]{http://dx.doi.org/}%
\providecommand \selectlanguage [0]{\@gobble}%
\providecommand \bibinfo  [0]{\@secondoftwo}%
\providecommand \bibfield  [0]{\@secondoftwo}%
\providecommand \translation [1]{[#1]}%
\providecommand \BibitemOpen [0]{}%
\providecommand \bibitemStop [0]{}%
\providecommand \bibitemNoStop [0]{.\EOS\space}%
\providecommand \EOS [0]{\spacefactor3000\relax}%
\providecommand \BibitemShut  [1]{\csname bibitem#1\endcsname}%
\let\auto@bib@innerbib\@empty
\bibitem [{\citenamefont {Cao}\ \emph {et~al.}(2018{\natexlab{a}})\citenamefont
  {Cao}, \citenamefont {Fatemi}, \citenamefont {Fang}, \citenamefont
  {Watanabe}, \citenamefont {Taniguchi}, \citenamefont {Kaxiras},\ and\
  \citenamefont {Jarillo-Herrero}}]{Cao2018a}%
  \BibitemOpen
  \bibfield  {author} {\bibinfo {author} {\bibfnamefont {Yuan}\ \bibnamefont
  {Cao}}, \bibinfo {author} {\bibfnamefont {Valla}\ \bibnamefont {Fatemi}},
  \bibinfo {author} {\bibfnamefont {Shiang}\ \bibnamefont {Fang}}, \bibinfo
  {author} {\bibfnamefont {Kenji}\ \bibnamefont {Watanabe}}, \bibinfo {author}
  {\bibfnamefont {Takashi}\ \bibnamefont {Taniguchi}}, \bibinfo {author}
  {\bibfnamefont {Efthimios}\ \bibnamefont {Kaxiras}}, \ and\ \bibinfo {author}
  {\bibfnamefont {Pablo}\ \bibnamefont {Jarillo-Herrero}},\ }\bibfield  {title}
  {\enquote {\bibinfo {title} {Unconventional superconductivity in magic-angle
  graphene superlattices},}\ }\href {http://dx.doi.org/10.1038/nature26160}
  {\bibfield  {journal} {\bibinfo  {journal} {Nature}\ ,\ \bibinfo {pages} {EP
  --}} (\bibinfo {year} {2018}{\natexlab{a}})}\BibitemShut {NoStop}%
\bibitem [{\citenamefont {Cao}\ \emph {et~al.}(2018{\natexlab{b}})\citenamefont
  {Cao}, \citenamefont {Fatemi}, \citenamefont {Demir}, \citenamefont {Fang},
  \citenamefont {Tomarken}, \citenamefont {Luo}, \citenamefont
  {Sanchez-Yamagishi}, \citenamefont {Watanabe}, \citenamefont {Taniguchi},
  \citenamefont {Kaxiras}, \citenamefont {Ashoori},\ and\ \citenamefont
  {Jarillo-Herrero}}]{Cao2018}%
  \BibitemOpen
  \bibfield  {author} {\bibinfo {author} {\bibfnamefont {Yuan}\ \bibnamefont
  {Cao}}, \bibinfo {author} {\bibfnamefont {Valla}\ \bibnamefont {Fatemi}},
  \bibinfo {author} {\bibfnamefont {Ahmet}\ \bibnamefont {Demir}}, \bibinfo
  {author} {\bibfnamefont {Shiang}\ \bibnamefont {Fang}}, \bibinfo {author}
  {\bibfnamefont {Spencer~L.}\ \bibnamefont {Tomarken}}, \bibinfo {author}
  {\bibfnamefont {Jason~Y.}\ \bibnamefont {Luo}}, \bibinfo {author}
  {\bibfnamefont {J.~D.}\ \bibnamefont {Sanchez-Yamagishi}}, \bibinfo {author}
  {\bibfnamefont {K.}~\bibnamefont {Watanabe}}, \bibinfo {author}
  {\bibfnamefont {T.}~\bibnamefont {Taniguchi}}, \bibinfo {author}
  {\bibfnamefont {E.}~\bibnamefont {Kaxiras}}, \bibinfo {author} {\bibfnamefont
  {R.~C.}\ \bibnamefont {Ashoori}}, \ and\ \bibinfo {author} {\bibfnamefont
  {P.}~\bibnamefont {Jarillo-Herrero}},\ }\bibfield  {title} {\enquote
  {\bibinfo {title} {Correlated insulator behaviour at half-filling in
  magic-angle graphene superlattices},}\ }\href
  {http://dx.doi.org/10.1038/nature26154} {\bibfield  {journal} {\bibinfo
  {journal} {Nature}\ ,\ \bibinfo {pages} {EP --}} (\bibinfo {year}
  {2018}{\natexlab{b}})}\BibitemShut {NoStop}%
\bibitem [{\citenamefont {Volovik}(2018)}]{Volovik2018}%
  \BibitemOpen
  \bibfield  {author} {\bibinfo {author} {\bibfnamefont {G.~E.}\ \bibnamefont
  {Volovik}},\ }\bibfield  {title} {\enquote {\bibinfo {title} {Graphite,
  graphene and the flat band superconductivity},}\ }\href {\doibase
  10.1134/S0021364018080052} {\bibfield  {journal} {\bibinfo  {journal} {JETP
  Letters}\ } (\bibinfo {year} {2018}),\ 10.1134/S0021364018080052}\BibitemShut
  {NoStop}%
\bibitem [{\citenamefont {{Xu}}\ and\ \citenamefont
  {{Balents}}(2018)}]{Xu2018}%
  \BibitemOpen
  \bibfield  {author} {\bibinfo {author} {\bibfnamefont {C.}~\bibnamefont
  {{Xu}}}\ and\ \bibinfo {author} {\bibfnamefont {L.}~\bibnamefont
  {{Balents}}},\ }\bibfield  {title} {\enquote {\bibinfo {title} {{Topological
  Superconductivity in Twisted Multilayer Graphene}},}\ }\href@noop {}
  {\bibfield  {journal} {\bibinfo  {journal} {ArXiv e-prints}\ } (\bibinfo
  {year} {2018})},\ \Eprint {http://arxiv.org/abs/1803.08057}
  {arXiv:1803.08057} \BibitemShut {NoStop}%
\bibitem [{\citenamefont {{Yuan}}\ and\ \citenamefont
  {{Fu}}(2018)}]{NoahLiang}%
  \BibitemOpen
  \bibfield  {author} {\bibinfo {author} {\bibfnamefont {N.~F.~Q.}\
  \bibnamefont {{Yuan}}}\ and\ \bibinfo {author} {\bibfnamefont
  {L.}~\bibnamefont {{Fu}}},\ }\bibfield  {title} {\enquote {\bibinfo {title}
  {{A Model for Metal-Insulator Transition in Graphene Superlattices and
  Beyond}},}\ }\href@noop {} {\bibfield  {journal} {\bibinfo  {journal} {ArXiv
  e-prints}\ } (\bibinfo {year} {2018})},\ \Eprint
  {http://arxiv.org/abs/1803.09699} {arXiv:1803.09699} \BibitemShut {NoStop}%
\bibitem [{\citenamefont {{Po}}\ \emph {et~al.}(2018)\citenamefont {{Po}},
  \citenamefont {{Zou}}, \citenamefont {{Vishwanath}},\ and\ \citenamefont
  {{Senthil}}}]{Mar2018}%
  \BibitemOpen
  \bibfield  {author} {\bibinfo {author} {\bibfnamefont {H.~C.}\ \bibnamefont
  {{Po}}}, \bibinfo {author} {\bibfnamefont {L.}~\bibnamefont {{Zou}}},
  \bibinfo {author} {\bibfnamefont {A.}~\bibnamefont {{Vishwanath}}}, \ and\
  \bibinfo {author} {\bibfnamefont {T.}~\bibnamefont {{Senthil}}},\ }\bibfield
  {title} {\enquote {\bibinfo {title} {{Origin of Mott insulating behavior and
  superconductivity in twisted bilayer graphene}},}\ }\href@noop {} {\bibfield
  {journal} {\bibinfo  {journal} {ArXiv e-prints}\ } (\bibinfo {year}
  {2018})},\ \Eprint {http://arxiv.org/abs/1803.09742} {arXiv:1803.09742}
  \BibitemShut {NoStop}%
\bibitem [{\citenamefont {{Roy}}\ and\ \citenamefont
  {{Juricic}}(2018)}]{Roy2018}%
  \BibitemOpen
  \bibfield  {author} {\bibinfo {author} {\bibfnamefont {B.}~\bibnamefont
  {{Roy}}}\ and\ \bibinfo {author} {\bibfnamefont {V.}~\bibnamefont
  {{Juricic}}},\ }\bibfield  {title} {\enquote {\bibinfo {title}
  {{Unconventional superconductivity in nearly flat bands in twisted bilayer
  graphene}},}\ }\href@noop {} {\bibfield  {journal} {\bibinfo  {journal}
  {ArXiv e-prints}\ } (\bibinfo {year} {2018})},\ \Eprint
  {http://arxiv.org/abs/1803.11190} {arXiv:1803.11190 [cond-mat.mes-hall]}
  \BibitemShut {NoStop}%
\bibitem [{\citenamefont {{Guo}}\ \emph {et~al.}(2018)\citenamefont {{Guo}},
  \citenamefont {{Zhu}}, \citenamefont {{Feng}},\ and\ \citenamefont
  {{Scalettar}}}]{Guo2018}%
  \BibitemOpen
  \bibfield  {author} {\bibinfo {author} {\bibfnamefont {H.}~\bibnamefont
  {{Guo}}}, \bibinfo {author} {\bibfnamefont {X.}~\bibnamefont {{Zhu}}},
  \bibinfo {author} {\bibfnamefont {S.}~\bibnamefont {{Feng}}}, \ and\ \bibinfo
  {author} {\bibfnamefont {R.~T.}\ \bibnamefont {{Scalettar}}},\ }\bibfield
  {title} {\enquote {\bibinfo {title} {{Pairing symmetry of interacting
  fermions on twisted bilayer graphene superlattice}},}\ }\href@noop {}
  {\bibfield  {journal} {\bibinfo  {journal} {ArXiv e-prints}\ } (\bibinfo
  {year} {2018})},\ \Eprint {http://arxiv.org/abs/1804.00159} {arXiv:1804.00159
  [cond-mat.str-el]} \BibitemShut {NoStop}%
\bibitem [{\citenamefont {{Baskaran}}(2018)}]{Baskaran2018}%
  \BibitemOpen
  \bibfield  {author} {\bibinfo {author} {\bibfnamefont {G.}~\bibnamefont
  {{Baskaran}}},\ }\bibfield  {title} {\enquote {\bibinfo {title} {{Theory of
  Emergent Josephson Lattice in Neutral Twisted Bilayer Graphene
  (Moi$\backslash$'re is Different)}},}\ }\href@noop {} {\bibfield  {journal}
  {\bibinfo  {journal} {ArXiv e-prints}\ } (\bibinfo {year} {2018})},\ \Eprint
  {http://arxiv.org/abs/1804.00627} {arXiv:1804.00627 [cond-mat.supr-con]}
  \BibitemShut {NoStop}%
\bibitem [{\citenamefont {{Padhi}}\ \emph {et~al.}(2018)\citenamefont
  {{Padhi}}, \citenamefont {{Setty}},\ and\ \citenamefont
  {{Phillips}}}]{Padhi2018}%
  \BibitemOpen
  \bibfield  {author} {\bibinfo {author} {\bibfnamefont {B.}~\bibnamefont
  {{Padhi}}}, \bibinfo {author} {\bibfnamefont {C.}~\bibnamefont {{Setty}}}, \
  and\ \bibinfo {author} {\bibfnamefont {P.~W.}\ \bibnamefont {{Phillips}}},\
  }\bibfield  {title} {\enquote {\bibinfo {title} {{Wigner Crystallization in
  lieu of Mottness in Twisted Bilayer Graphene}},}\ }\href@noop {} {\bibfield
  {journal} {\bibinfo  {journal} {ArXiv e-prints}\ } (\bibinfo {year}
  {2018})},\ \Eprint {http://arxiv.org/abs/1804.01101} {arXiv:1804.01101
  [cond-mat.str-el]} \BibitemShut {NoStop}%
\bibitem [{\citenamefont {{Irkhin}}\ and\ \citenamefont
  {{Skryabin}}(2018)}]{Irkhin2018}%
  \BibitemOpen
  \bibfield  {author} {\bibinfo {author} {\bibfnamefont {V.~Y.}\ \bibnamefont
  {{Irkhin}}}\ and\ \bibinfo {author} {\bibfnamefont {Y.~N.}\ \bibnamefont
  {{Skryabin}}},\ }\bibfield  {title} {\enquote {\bibinfo {title} {{Dirac
  points, spinons and spin liquid in twisted bilayer graphene}},}\ }\href
  {\doibase 10.1134/S0021364018100016} {\bibfield  {journal} {\bibinfo
  {journal} {Soviet Journal of Experimental and Theoretical Physics Letters}\ }
  (\bibinfo {year} {2018}),\ 10.1134/S0021364018100016},\ \Eprint
  {http://arxiv.org/abs/1804.02236} {arXiv:1804.02236 [cond-mat.str-el]}
  \BibitemShut {NoStop}%
\bibitem [{\citenamefont {{Dodaro}}\ \emph {et~al.}(2018)\citenamefont
  {{Dodaro}}, \citenamefont {{Kivelson}}, \citenamefont {{Schattner}},
  \citenamefont {{Sun}},\ and\ \citenamefont {{Wang}}}]{Dodaro2018}%
  \BibitemOpen
  \bibfield  {author} {\bibinfo {author} {\bibfnamefont {J.~F.}\ \bibnamefont
  {{Dodaro}}}, \bibinfo {author} {\bibfnamefont {S.~A.}\ \bibnamefont
  {{Kivelson}}}, \bibinfo {author} {\bibfnamefont {Y.}~\bibnamefont
  {{Schattner}}}, \bibinfo {author} {\bibfnamefont {X.-Q.}\ \bibnamefont
  {{Sun}}}, \ and\ \bibinfo {author} {\bibfnamefont {C.}~\bibnamefont
  {{Wang}}},\ }\bibfield  {title} {\enquote {\bibinfo {title} {{Phases of a
  phenomenological model of twisted bilayer graphene}},}\ }\href@noop {}
  {\bibfield  {journal} {\bibinfo  {journal} {ArXiv e-prints}\ } (\bibinfo
  {year} {2018})},\ \Eprint {http://arxiv.org/abs/1804.03162} {arXiv:1804.03162
  [cond-mat.supr-con]} \BibitemShut {NoStop}%
\bibitem [{\citenamefont {{Huang}}\ \emph {et~al.}(2018)\citenamefont
  {{Huang}}, \citenamefont {{Zhang}},\ and\ \citenamefont {{Ma}}}]{Huang2018}%
  \BibitemOpen
  \bibfield  {author} {\bibinfo {author} {\bibfnamefont {T.}~\bibnamefont
  {{Huang}}}, \bibinfo {author} {\bibfnamefont {L.}~\bibnamefont {{Zhang}}}, \
  and\ \bibinfo {author} {\bibfnamefont {T.}~\bibnamefont {{Ma}}},\ }\bibfield
  {title} {\enquote {\bibinfo {title} {{Antiferromagnetically ordered Mott
  insulator and $d+id$ superconductivity in twisted bilayer graphene: A quantum
  Monte carlo study}},}\ }\href@noop {} {\bibfield  {journal} {\bibinfo
  {journal} {ArXiv e-prints}\ } (\bibinfo {year} {2018})},\ \Eprint
  {http://arxiv.org/abs/1804.06096} {arXiv:1804.06096 [cond-mat.supr-con]}
  \BibitemShut {NoStop}%
\bibitem [{\citenamefont {{Zhang}}(2018)}]{Zhang2018}%
  \BibitemOpen
  \bibfield  {author} {\bibinfo {author} {\bibfnamefont {L.}~\bibnamefont
  {{Zhang}}},\ }\bibfield  {title} {\enquote {\bibinfo {title} {{Low-energy
  Moir$\backslash$'e Band Formed by Dirac Zero Modes in Twisted Bilayer
  Graphene}},}\ }\href@noop {} {\bibfield  {journal} {\bibinfo  {journal}
  {ArXiv e-prints}\ } (\bibinfo {year} {2018})},\ \Eprint
  {http://arxiv.org/abs/1804.09047} {arXiv:1804.09047 [cond-mat.mes-hall]}
  \BibitemShut {NoStop}%
\bibitem [{\citenamefont {{Ray}}\ and\ \citenamefont {{Das}}(2018)}]{Ray2018}%
  \BibitemOpen
  \bibfield  {author} {\bibinfo {author} {\bibfnamefont {S.}~\bibnamefont
  {{Ray}}}\ and\ \bibinfo {author} {\bibfnamefont {T.}~\bibnamefont {{Das}}},\
  }\bibfield  {title} {\enquote {\bibinfo {title} {{Wannier Pairs in the
  Superconducting Twisted Bilayer Graphene and Related Systems}},}\ }\href@noop
  {} {\bibfield  {journal} {\bibinfo  {journal} {ArXiv e-prints}\ } (\bibinfo
  {year} {2018})},\ \Eprint {http://arxiv.org/abs/1804.09674} {arXiv:1804.09674
  [cond-mat.supr-con]} \BibitemShut {NoStop}%
\bibitem [{\citenamefont {{Liu}}\ \emph {et~al.}(2018)\citenamefont {{Liu}},
  \citenamefont {{Zhang}}, \citenamefont {{Chen}},\ and\ \citenamefont
  {{Yang}}}]{Liu2018}%
  \BibitemOpen
  \bibfield  {author} {\bibinfo {author} {\bibfnamefont {C.-C.}\ \bibnamefont
  {{Liu}}}, \bibinfo {author} {\bibfnamefont {L.-D.}\ \bibnamefont {{Zhang}}},
  \bibinfo {author} {\bibfnamefont {W.-Q.}\ \bibnamefont {{Chen}}}, \ and\
  \bibinfo {author} {\bibfnamefont {F.}~\bibnamefont {{Yang}}},\ }\bibfield
  {title} {\enquote {\bibinfo {title} {{Chiral SDW and d + id superconductivity
  in the magic-angle twisted bilayer-graphene}},}\ }\href@noop {} {\bibfield
  {journal} {\bibinfo  {journal} {ArXiv e-prints}\ } (\bibinfo {year}
  {2018})},\ \Eprint {http://arxiv.org/abs/1804.10009} {arXiv:1804.10009
  [cond-mat.supr-con]} \BibitemShut {NoStop}%
\bibitem [{\citenamefont {{Xu}}\ \emph {et~al.}(2018)\citenamefont {{Xu}},
  \citenamefont {{Law}},\ and\ \citenamefont {{Lee}}}]{XuLawLee2018}%
  \BibitemOpen
  \bibfield  {author} {\bibinfo {author} {\bibfnamefont {X.~Y.}\ \bibnamefont
  {{Xu}}}, \bibinfo {author} {\bibfnamefont {K.~T.}\ \bibnamefont {{Law}}}, \
  and\ \bibinfo {author} {\bibfnamefont {P.~A.}\ \bibnamefont {{Lee}}},\
  }\bibfield  {title} {\enquote {\bibinfo {title} {{Kekul$\backslash$'e valence
  bond order in an extended Hubbard model on the honeycomb lattice, with
  possible applications to twisted bilayer graphene}},}\ }\href@noop {}
  {\bibfield  {journal} {\bibinfo  {journal} {ArXiv e-prints}\ } (\bibinfo
  {year} {2018})},\ \Eprint {http://arxiv.org/abs/1805.00478} {arXiv:1805.00478
  [cond-mat.str-el]} \BibitemShut {NoStop}%
\bibitem [{\citenamefont {{Kang}}\ and\ \citenamefont {{Vafek}}(2018)}]{Oskar}%
  \BibitemOpen
  \bibfield  {author} {\bibinfo {author} {\bibfnamefont {J.}~\bibnamefont
  {{Kang}}}\ and\ \bibinfo {author} {\bibfnamefont {O.}~\bibnamefont
  {{Vafek}}},\ }\bibfield  {title} {\enquote {\bibinfo {title} {{Symmetry,
  maximally localized Wannier states, and low energy model for the twisted
  bilayer graphene narrow bands}},}\ }\href@noop {} {\bibfield  {journal}
  {\bibinfo  {journal} {ArXiv e-prints}\ } (\bibinfo {year} {2018})},\ \Eprint
  {http://arxiv.org/abs/1805.04918} {arXiv:1805.04918} \BibitemShut {NoStop}%
\bibitem [{\citenamefont {{Rademaker}}\ and\ \citenamefont
  {{Mellado}}(2018)}]{Rademaker2018}%
  \BibitemOpen
  \bibfield  {author} {\bibinfo {author} {\bibfnamefont {L.}~\bibnamefont
  {{Rademaker}}}\ and\ \bibinfo {author} {\bibfnamefont {P.}~\bibnamefont
  {{Mellado}}},\ }\bibfield  {title} {\enquote {\bibinfo {title}
  {{Charge-transfer insulation in twisted bilayer graphene}},}\ }\href@noop {}
  {\bibfield  {journal} {\bibinfo  {journal} {ArXiv e-prints}\ } (\bibinfo
  {year} {2018})},\ \Eprint {http://arxiv.org/abs/1805.05294} {arXiv:1805.05294
  [cond-mat.str-el]} \BibitemShut {NoStop}%
\bibitem [{\citenamefont {{Isobe}}\ \emph {et~al.}(2018)\citenamefont
  {{Isobe}}, \citenamefont {{Yuan}},\ and\ \citenamefont {{Fu}}}]{Isobe2018}%
  \BibitemOpen
  \bibfield  {author} {\bibinfo {author} {\bibfnamefont {H.}~\bibnamefont
  {{Isobe}}}, \bibinfo {author} {\bibfnamefont {N.~F.~Q.}\ \bibnamefont
  {{Yuan}}}, \ and\ \bibinfo {author} {\bibfnamefont {L.}~\bibnamefont
  {{Fu}}},\ }\bibfield  {title} {\enquote {\bibinfo {title} {{Superconductivity
  and Charge Density Wave in Twisted Bilayer Graphene}},}\ }\href@noop {}
  {\bibfield  {journal} {\bibinfo  {journal} {ArXiv e-prints}\ } (\bibinfo
  {year} {2018})},\ \Eprint {http://arxiv.org/abs/1805.06449} {arXiv:1805.06449
  [cond-mat.str-el]} \BibitemShut {NoStop}%
\bibitem [{\citenamefont {{Koshino}}\ \emph {et~al.}(2018)\citenamefont
  {{Koshino}}, \citenamefont {{Yuan}}, \citenamefont {{Ochi}}, \citenamefont
  {{Kuroki}},\ and\ \citenamefont {{Fu}}}]{KoshinoLiang}%
  \BibitemOpen
  \bibfield  {author} {\bibinfo {author} {\bibfnamefont {M.}~\bibnamefont
  {{Koshino}}}, \bibinfo {author} {\bibfnamefont {N.~F.~Q.}\ \bibnamefont
  {{Yuan}}}, \bibinfo {author} {\bibfnamefont {M.}~\bibnamefont {{Ochi}}},
  \bibinfo {author} {\bibfnamefont {K.}~\bibnamefont {{Kuroki}}}, \ and\
  \bibinfo {author} {\bibfnamefont {L.}~\bibnamefont {{Fu}}},\ }\bibfield
  {title} {\enquote {\bibinfo {title} {{Maximally-localized Wannier orbitals
  and the extended Hubbard model for the twisted bilayer graphene}},}\
  }\href@noop {} {\bibfield  {journal} {\bibinfo  {journal} {ArXiv e-prints}\ }
  (\bibinfo {year} {2018})},\ \Eprint {http://arxiv.org/abs/1805.06819}
  {arXiv:1805.06819} \BibitemShut {NoStop}%
\bibitem [{\citenamefont {{Wu}}\ \emph
  {et~al.}(2018{\natexlab{a}})\citenamefont {{Wu}}, \citenamefont
  {{MacDonald}},\ and\ \citenamefont {{Martin}}}]{Wu2018}%
  \BibitemOpen
  \bibfield  {author} {\bibinfo {author} {\bibfnamefont {F.}~\bibnamefont
  {{Wu}}}, \bibinfo {author} {\bibfnamefont {A.~H.}\ \bibnamefont
  {{MacDonald}}}, \ and\ \bibinfo {author} {\bibfnamefont {I.}~\bibnamefont
  {{Martin}}},\ }\bibfield  {title} {\enquote {\bibinfo {title} {{Theory of
  phonon-mediated superconductivity in twisted bilayer graphene}},}\
  }\href@noop {} {\bibfield  {journal} {\bibinfo  {journal} {ArXiv e-prints}\ }
  (\bibinfo {year} {2018}{\natexlab{a}})},\ \Eprint
  {http://arxiv.org/abs/1805.08735} {arXiv:1805.08735 [cond-mat.supr-con]}
  \BibitemShut {NoStop}%
\bibitem [{\citenamefont {{Pizarro}}\ \emph {et~al.}(2018)\citenamefont
  {{Pizarro}}, \citenamefont {{Calder{\'o}n}},\ and\ \citenamefont
  {{Bascones}}}]{Pizarro2018}%
  \BibitemOpen
  \bibfield  {author} {\bibinfo {author} {\bibfnamefont {J.~M.}\ \bibnamefont
  {{Pizarro}}}, \bibinfo {author} {\bibfnamefont {M.~J.}\ \bibnamefont
  {{Calder{\'o}n}}}, \ and\ \bibinfo {author} {\bibfnamefont {E.}~\bibnamefont
  {{Bascones}}},\ }\bibfield  {title} {\enquote {\bibinfo {title} {{The nature
  of correlations in the insulating states of twisted bilayer graphene}},}\
  }\href@noop {} {\bibfield  {journal} {\bibinfo  {journal} {ArXiv e-prints}\ }
  (\bibinfo {year} {2018})},\ \Eprint {http://arxiv.org/abs/1805.07303}
  {arXiv:1805.07303 [cond-mat.str-el]} \BibitemShut {NoStop}%
\bibitem [{\citenamefont {{Peltonen}}\ \emph {et~al.}(2018)\citenamefont
  {{Peltonen}}, \citenamefont {{Ojaj{\"a}rvi}},\ and\ \citenamefont
  {{Heikkil{\"a}}}}]{Peltonen2018}%
  \BibitemOpen
  \bibfield  {author} {\bibinfo {author} {\bibfnamefont {T.~J.}\ \bibnamefont
  {{Peltonen}}}, \bibinfo {author} {\bibfnamefont {R.}~\bibnamefont
  {{Ojaj{\"a}rvi}}}, \ and\ \bibinfo {author} {\bibfnamefont {T.~T.}\
  \bibnamefont {{Heikkil{\"a}}}},\ }\bibfield  {title} {\enquote {\bibinfo
  {title} {{Mean-field theory for superconductivity in twisted bilayer
  graphene}},}\ }\href@noop {} {\bibfield  {journal} {\bibinfo  {journal}
  {ArXiv e-prints}\ } (\bibinfo {year} {2018})},\ \Eprint
  {http://arxiv.org/abs/1805.01039} {arXiv:1805.01039 [cond-mat.supr-con]}
  \BibitemShut {NoStop}%
\bibitem [{\citenamefont {{You}}\ and\ \citenamefont
  {{Vishwanath}}(2018)}]{You2018}%
  \BibitemOpen
  \bibfield  {author} {\bibinfo {author} {\bibfnamefont {Y.-Z.}\ \bibnamefont
  {{You}}}\ and\ \bibinfo {author} {\bibfnamefont {A.}~\bibnamefont
  {{Vishwanath}}},\ }\bibfield  {title} {\enquote {\bibinfo {title}
  {{Superconductivity from Valley Fluctuations and Approximate SO(4) Symmetry
  in a Weak Coupling Theory of Twisted Bilayer Graphene}},}\ }\href@noop {} {\
  (\bibinfo {year} {2018})},\ \Eprint {http://arxiv.org/abs/1805.06867}
  {arXiv:1805.06867} \BibitemShut {NoStop}%
\bibitem [{\citenamefont {{Wu}}\ \emph
  {et~al.}(2018{\natexlab{b}})\citenamefont {{Wu}}, \citenamefont {{Pawlak}},
  \citenamefont {{Jian}},\ and\ \citenamefont {{Xu}}}]{WuXu2018}%
  \BibitemOpen
  \bibfield  {author} {\bibinfo {author} {\bibfnamefont {X.-C.}\ \bibnamefont
  {{Wu}}}, \bibinfo {author} {\bibfnamefont {K.~A.}\ \bibnamefont {{Pawlak}}},
  \bibinfo {author} {\bibfnamefont {C.-M.}\ \bibnamefont {{Jian}}}, \ and\
  \bibinfo {author} {\bibfnamefont {C.}~\bibnamefont {{Xu}}},\ }\bibfield
  {title} {\enquote {\bibinfo {title} {{Emergent Superconductivity in the weak
  Mott insulator phase of bilayer Graphene Moir$\backslash$'e Superlattice}},}\
  }\href@noop {} {\bibfield  {journal} {\bibinfo  {journal} {ArXiv e-prints}\ }
  (\bibinfo {year} {2018}{\natexlab{b}})},\ \Eprint
  {http://arxiv.org/abs/1805.06906} {arXiv:1805.06906 [cond-mat.str-el]}
  \BibitemShut {NoStop}%
\bibitem [{\citenamefont {{Pal}}(2018)}]{Pal2018}%
  \BibitemOpen
  \bibfield  {author} {\bibinfo {author} {\bibfnamefont {H.~K.}\ \bibnamefont
  {{Pal}}},\ }\bibfield  {title} {\enquote {\bibinfo {title} {{On magic angles
  and band flattening in twisted bilayer graphene}},}\ }\href@noop {}
  {\bibfield  {journal} {\bibinfo  {journal} {ArXiv e-prints}\ } (\bibinfo
  {year} {2018})},\ \Eprint {http://arxiv.org/abs/1805.08803} {arXiv:1805.08803
  [cond-mat.mes-hall]} \BibitemShut {NoStop}%
\bibitem [{\citenamefont {{Ochi}}\ \emph {et~al.}(2018)\citenamefont {{Ochi}},
  \citenamefont {{Koshino}},\ and\ \citenamefont {{Kuroki}}}]{Ochi2018}%
  \BibitemOpen
  \bibfield  {author} {\bibinfo {author} {\bibfnamefont {M.}~\bibnamefont
  {{Ochi}}}, \bibinfo {author} {\bibfnamefont {M.}~\bibnamefont {{Koshino}}}, \
  and\ \bibinfo {author} {\bibfnamefont {K.}~\bibnamefont {{Kuroki}}},\
  }\bibfield  {title} {\enquote {\bibinfo {title} {{Possible correlated
  insulating states in magic-angle twisted bilayer graphene under strongly
  competing interactions}},}\ }\href@noop {} {\bibfield  {journal} {\bibinfo
  {journal} {ArXiv e-prints}\ } (\bibinfo {year} {2018})},\ \Eprint
  {http://arxiv.org/abs/1805.09606} {arXiv:1805.09606 [cond-mat.str-el]}
  \BibitemShut {NoStop}%
\bibitem [{\citenamefont {{Fidrysiak}}\ \emph {et~al.}(2018)\citenamefont
  {{Fidrysiak}}, \citenamefont {{Zegrodnik}},\ and\ \citenamefont
  {{Spa{\l}ek}}}]{Fidrysiak2018}%
  \BibitemOpen
  \bibfield  {author} {\bibinfo {author} {\bibfnamefont {M.}~\bibnamefont
  {{Fidrysiak}}}, \bibinfo {author} {\bibfnamefont {M.}~\bibnamefont
  {{Zegrodnik}}}, \ and\ \bibinfo {author} {\bibfnamefont {J.}~\bibnamefont
  {{Spa{\l}ek}}},\ }\bibfield  {title} {\enquote {\bibinfo {title}
  {{Unconventional topological superconductivity and phase diagram for a
  two-orbital model of twisted bilayer graphene}},}\ }\href@noop {} {\bibfield
  {journal} {\bibinfo  {journal} {ArXiv e-prints}\ } (\bibinfo {year}
  {2018})},\ \Eprint {http://arxiv.org/abs/1805.01179} {arXiv:1805.01179
  [cond-mat.supr-con]} \BibitemShut {NoStop}%
\bibitem [{\citenamefont {{Thomson}}\ \emph {et~al.}(2018)\citenamefont
  {{Thomson}}, \citenamefont {{Chatterjee}}, \citenamefont {{Sachdev}},\ and\
  \citenamefont {{Scheurer}}}]{Thomson2018}%
  \BibitemOpen
  \bibfield  {author} {\bibinfo {author} {\bibfnamefont {A.}~\bibnamefont
  {{Thomson}}}, \bibinfo {author} {\bibfnamefont {S.}~\bibnamefont
  {{Chatterjee}}}, \bibinfo {author} {\bibfnamefont {S.}~\bibnamefont
  {{Sachdev}}}, \ and\ \bibinfo {author} {\bibfnamefont {M.~S.}\ \bibnamefont
  {{Scheurer}}},\ }\bibfield  {title} {\enquote {\bibinfo {title} {{Triangular
  antiferromagnetism on the honeycomb lattice of twisted bilayer graphene}},}\
  }\href@noop {} {\bibfield  {journal} {\bibinfo  {journal} {ArXiv e-prints}\ }
  (\bibinfo {year} {2018})},\ \Eprint {http://arxiv.org/abs/1806.02837}
  {arXiv:1806.02837 [cond-mat.str-el]} \BibitemShut {NoStop}%
\bibitem [{\citenamefont {{Guinea}}\ and\ \citenamefont
  {{Walet}}(2018)}]{Guinea2018}%
  \BibitemOpen
  \bibfield  {author} {\bibinfo {author} {\bibfnamefont {F.}~\bibnamefont
  {{Guinea}}}\ and\ \bibinfo {author} {\bibfnamefont {N.~R.}\ \bibnamefont
  {{Walet}}},\ }\bibfield  {title} {\enquote {\bibinfo {title} {{Electrostatic
  effects and band distortions in twisted graphene bilayers}},}\ }\href@noop {}
  {\bibfield  {journal} {\bibinfo  {journal} {ArXiv e-prints}\ } (\bibinfo
  {year} {2018})},\ \Eprint {http://arxiv.org/abs/1806.05990} {arXiv:1806.05990
  [cond-mat.mes-hall]} \BibitemShut {NoStop}%
\bibitem [{\citenamefont {Lopes~dos Santos}\ \emph {et~al.}(2007)\citenamefont
  {Lopes~dos Santos}, \citenamefont {Peres},\ and\ \citenamefont
  {Castro~Neto}}]{Neto2007}%
  \BibitemOpen
  \bibfield  {author} {\bibinfo {author} {\bibfnamefont {J.~M.~B.}\
  \bibnamefont {Lopes~dos Santos}}, \bibinfo {author} {\bibfnamefont
  {N.~M.~R.}\ \bibnamefont {Peres}}, \ and\ \bibinfo {author} {\bibfnamefont
  {A.~H.}\ \bibnamefont {Castro~Neto}},\ }\bibfield  {title} {\enquote
  {\bibinfo {title} {Graphene bilayer with a twist: Electronic structure},}\
  }\href {\doibase 10.1103/PhysRevLett.99.256802} {\bibfield  {journal}
  {\bibinfo  {journal} {Phys. Rev. Lett.}\ }\textbf {\bibinfo {volume} {99}},\
  \bibinfo {pages} {256802} (\bibinfo {year} {2007})}\BibitemShut {NoStop}%
\bibitem [{\citenamefont {Bistritzer}\ and\ \citenamefont
  {MacDonald}(2011)}]{Bistritzer2011}%
  \BibitemOpen
  \bibfield  {author} {\bibinfo {author} {\bibfnamefont {Rafi}\ \bibnamefont
  {Bistritzer}}\ and\ \bibinfo {author} {\bibfnamefont {Allan~H.}\ \bibnamefont
  {MacDonald}},\ }\bibfield  {title} {\enquote {\bibinfo {title} {Moir{\'e}
  bands in twisted double-layer graphene},}\ }\href {\doibase
  10.1073/pnas.1108174108} {\bibfield  {journal} {\bibinfo  {journal}
  {Proceedings of the National Academy of Sciences}\ }\textbf {\bibinfo
  {volume} {108}},\ \bibinfo {pages} {12233--12237} (\bibinfo {year} {2011})},\
  \Eprint
  {http://arxiv.org/abs/http://www.pnas.org/content/108/30/12233.full.pdf}
  {http://www.pnas.org/content/108/30/12233.full.pdf} \BibitemShut {NoStop}%
\bibitem [{\citenamefont {Shallcross}\ \emph {et~al.}(2010)\citenamefont
  {Shallcross}, \citenamefont {Sharma}, \citenamefont {Kandelaki},\ and\
  \citenamefont {Pankratov}}]{Shallcross}%
  \BibitemOpen
  \bibfield  {author} {\bibinfo {author} {\bibfnamefont {S.}~\bibnamefont
  {Shallcross}}, \bibinfo {author} {\bibfnamefont {S.}~\bibnamefont {Sharma}},
  \bibinfo {author} {\bibfnamefont {E.}~\bibnamefont {Kandelaki}}, \ and\
  \bibinfo {author} {\bibfnamefont {O.~A.}\ \bibnamefont {Pankratov}},\
  }\bibfield  {title} {\enquote {\bibinfo {title} {Electronic structure of
  turbostratic graphene},}\ }\href {\doibase 10.1103/PhysRevB.81.165105}
  {\bibfield  {journal} {\bibinfo  {journal} {Phys. Rev. B}\ }\textbf {\bibinfo
  {volume} {81}},\ \bibinfo {pages} {165105} (\bibinfo {year}
  {2010})}\BibitemShut {NoStop}%
\bibitem [{\citenamefont {Su\'arez~Morell}\ \emph {et~al.}(2010)\citenamefont
  {Su\'arez~Morell}, \citenamefont {Correa}, \citenamefont {Vargas},
  \citenamefont {Pacheco},\ and\ \citenamefont {Barticevic}}]{Morell2010}%
  \BibitemOpen
  \bibfield  {author} {\bibinfo {author} {\bibfnamefont {E.}~\bibnamefont
  {Su\'arez~Morell}}, \bibinfo {author} {\bibfnamefont {J.~D.}\ \bibnamefont
  {Correa}}, \bibinfo {author} {\bibfnamefont {P.}~\bibnamefont {Vargas}},
  \bibinfo {author} {\bibfnamefont {M.}~\bibnamefont {Pacheco}}, \ and\
  \bibinfo {author} {\bibfnamefont {Z.}~\bibnamefont {Barticevic}},\ }\bibfield
   {title} {\enquote {\bibinfo {title} {Flat bands in slightly twisted bilayer
  graphene: Tight-binding calculations},}\ }\href {\doibase
  10.1103/PhysRevB.82.121407} {\bibfield  {journal} {\bibinfo  {journal} {Phys.
  Rev. B}\ }\textbf {\bibinfo {volume} {82}},\ \bibinfo {pages} {121407}
  (\bibinfo {year} {2010})}\BibitemShut {NoStop}%
\bibitem [{\citenamefont {Trambly~de Laissardi{\`e}re}\ \emph
  {et~al.}(2010)\citenamefont {Trambly~de Laissardi{\`e}re}, \citenamefont
  {Mayou},\ and\ \citenamefont {Magaud}}]{Mayou2010}%
  \BibitemOpen
  \bibfield  {author} {\bibinfo {author} {\bibfnamefont {G.}~\bibnamefont
  {Trambly~de Laissardi{\`e}re}}, \bibinfo {author} {\bibfnamefont
  {D.}~\bibnamefont {Mayou}}, \ and\ \bibinfo {author} {\bibfnamefont
  {L.}~\bibnamefont {Magaud}},\ }\bibfield  {title} {\enquote {\bibinfo {title}
  {Localization of dirac electrons in rotated graphene bilayers},}\ }\href
  {\doibase 10.1021/nl902948m} {\bibfield  {journal} {\bibinfo  {journal} {Nano
  Letters}\ }\textbf {\bibinfo {volume} {10}},\ \bibinfo {pages} {804--808}
  (\bibinfo {year} {2010})},\ \bibinfo {note} {pMID: 20121163},\ \Eprint
  {http://arxiv.org/abs/https://doi.org/10.1021/nl902948m}
  {https://doi.org/10.1021/nl902948m} \BibitemShut {NoStop}%
\bibitem [{\citenamefont {Jung}\ \emph {et~al.}(2014)\citenamefont {Jung},
  \citenamefont {Raoux}, \citenamefont {Qiao},\ and\ \citenamefont
  {MacDonald}}]{Jung2014}%
  \BibitemOpen
  \bibfield  {author} {\bibinfo {author} {\bibfnamefont {Jeil}\ \bibnamefont
  {Jung}}, \bibinfo {author} {\bibfnamefont {Arnaud}\ \bibnamefont {Raoux}},
  \bibinfo {author} {\bibfnamefont {Zhenhua}\ \bibnamefont {Qiao}}, \ and\
  \bibinfo {author} {\bibfnamefont {A.~H.}\ \bibnamefont {MacDonald}},\
  }\bibfield  {title} {\enquote {\bibinfo {title} {Ab initio theory of
  moir{\'e} superlattice bands in layered two-dimensional materials},}\ }\href
  {\doibase 10.1103/PhysRevB.89.205414} {\bibfield  {journal} {\bibinfo
  {journal} {Phys. Rev. B}\ }\textbf {\bibinfo {volume} {89}},\ \bibinfo
  {pages} {205414} (\bibinfo {year} {2014})}\BibitemShut {NoStop}%
\bibitem [{\citenamefont {Cao}\ \emph {et~al.}(2016)\citenamefont {Cao},
  \citenamefont {Luo}, \citenamefont {Fatemi}, \citenamefont {Fang},
  \citenamefont {Sanchez-Yamagishi}, \citenamefont {Watanabe}, \citenamefont
  {Taniguchi}, \citenamefont {Kaxiras},\ and\ \citenamefont
  {Jarillo-Herrero}}]{PabloPRL}%
  \BibitemOpen
  \bibfield  {author} {\bibinfo {author} {\bibfnamefont {Y.}~\bibnamefont
  {Cao}}, \bibinfo {author} {\bibfnamefont {J.~Y.}\ \bibnamefont {Luo}},
  \bibinfo {author} {\bibfnamefont {V.}~\bibnamefont {Fatemi}}, \bibinfo
  {author} {\bibfnamefont {S.}~\bibnamefont {Fang}}, \bibinfo {author}
  {\bibfnamefont {J.~D.}\ \bibnamefont {Sanchez-Yamagishi}}, \bibinfo {author}
  {\bibfnamefont {K.}~\bibnamefont {Watanabe}}, \bibinfo {author}
  {\bibfnamefont {T.}~\bibnamefont {Taniguchi}}, \bibinfo {author}
  {\bibfnamefont {E.}~\bibnamefont {Kaxiras}}, \ and\ \bibinfo {author}
  {\bibfnamefont {P.}~\bibnamefont {Jarillo-Herrero}},\ }\bibfield  {title}
  {\enquote {\bibinfo {title} {Superlattice-induced insulating states and
  valley-protected orbits in twisted bilayer graphene},}\ }\href {\doibase
  10.1103/PhysRevLett.117.116804} {\bibfield  {journal} {\bibinfo  {journal}
  {Phys. Rev. Lett.}\ }\textbf {\bibinfo {volume} {117}},\ \bibinfo {pages}
  {116804} (\bibinfo {year} {2016})}\BibitemShut {NoStop}%
\bibitem [{\citenamefont {Lopes~dos Santos}\ \emph {et~al.}(2012)\citenamefont
  {Lopes~dos Santos}, \citenamefont {Peres},\ and\ \citenamefont
  {Castro~Neto}}]{Castro-Neto2012}%
  \BibitemOpen
  \bibfield  {author} {\bibinfo {author} {\bibfnamefont {J.~M.~B.}\
  \bibnamefont {Lopes~dos Santos}}, \bibinfo {author} {\bibfnamefont
  {N.~M.~R.}\ \bibnamefont {Peres}}, \ and\ \bibinfo {author} {\bibfnamefont
  {A.~H.}\ \bibnamefont {Castro~Neto}},\ }\bibfield  {title} {\enquote
  {\bibinfo {title} {Continuum model of the twisted graphene bilayer},}\ }\href
  {\doibase 10.1103/PhysRevB.86.155449} {\bibfield  {journal} {\bibinfo
  {journal} {Phys. Rev. B}\ }\textbf {\bibinfo {volume} {86}},\ \bibinfo
  {pages} {155449} (\bibinfo {year} {2012})}\BibitemShut {NoStop}%
\bibitem [{\citenamefont {Mele}(2010)}]{Mele2010}%
  \BibitemOpen
  \bibfield  {author} {\bibinfo {author} {\bibfnamefont {E.~J.}\ \bibnamefont
  {Mele}},\ }\bibfield  {title} {\enquote {\bibinfo {title} {Commensuration and
  interlayer coherence in twisted bilayer graphene},}\ }\href {\doibase
  10.1103/PhysRevB.81.161405} {\bibfield  {journal} {\bibinfo  {journal} {Phys.
  Rev. B}\ }\textbf {\bibinfo {volume} {81}},\ \bibinfo {pages} {161405}
  (\bibinfo {year} {2010})}\BibitemShut {NoStop}%
\bibitem [{\citenamefont {Li}\ \emph {et~al.}(2009)\citenamefont {Li},
  \citenamefont {Luican}, \citenamefont {Lopes~dos Santos}, \citenamefont
  {Castro~Neto}, \citenamefont {Reina}, \citenamefont {Kong},\ and\
  \citenamefont {Andrei}}]{STMNatPhy}%
  \BibitemOpen
  \bibfield  {author} {\bibinfo {author} {\bibfnamefont {Guohong}\ \bibnamefont
  {Li}}, \bibinfo {author} {\bibfnamefont {A.}~\bibnamefont {Luican}}, \bibinfo
  {author} {\bibfnamefont {J.~M.~B.}\ \bibnamefont {Lopes~dos Santos}},
  \bibinfo {author} {\bibfnamefont {A.~H.}\ \bibnamefont {Castro~Neto}},
  \bibinfo {author} {\bibfnamefont {A.}~\bibnamefont {Reina}}, \bibinfo
  {author} {\bibfnamefont {J.}~\bibnamefont {Kong}}, \ and\ \bibinfo {author}
  {\bibfnamefont {E.~Y.}\ \bibnamefont {Andrei}},\ }\bibfield  {title}
  {\enquote {\bibinfo {title} {Observation of van hove singularities in twisted
  graphene layers},}\ }\href {http://dx.doi.org/10.1038/nphys1463} {\bibfield
  {journal} {\bibinfo  {journal} {Nature Physics}\ }\textbf {\bibinfo {volume}
  {6}},\ \bibinfo {pages} {109 EP --} (\bibinfo {year} {2009})}\BibitemShut
  {NoStop}%
\bibitem [{\citenamefont {Luican}\ \emph {et~al.}(2011)\citenamefont {Luican},
  \citenamefont {Li}, \citenamefont {Reina}, \citenamefont {Kong},
  \citenamefont {Nair}, \citenamefont {Novoselov}, \citenamefont {Geim},\ and\
  \citenamefont {Andrei}}]{STMPRL}%
  \BibitemOpen
  \bibfield  {author} {\bibinfo {author} {\bibfnamefont {A.}~\bibnamefont
  {Luican}}, \bibinfo {author} {\bibfnamefont {Guohong}\ \bibnamefont {Li}},
  \bibinfo {author} {\bibfnamefont {A.}~\bibnamefont {Reina}}, \bibinfo
  {author} {\bibfnamefont {J.}~\bibnamefont {Kong}}, \bibinfo {author}
  {\bibfnamefont {R.~R.}\ \bibnamefont {Nair}}, \bibinfo {author}
  {\bibfnamefont {K.~S.}\ \bibnamefont {Novoselov}}, \bibinfo {author}
  {\bibfnamefont {A.~K.}\ \bibnamefont {Geim}}, \ and\ \bibinfo {author}
  {\bibfnamefont {E.~Y.}\ \bibnamefont {Andrei}},\ }\bibfield  {title}
  {\enquote {\bibinfo {title} {Single-layer behavior and its breakdown in
  twisted graphene layers},}\ }\href {\doibase 10.1103/PhysRevLett.106.126802}
  {\bibfield  {journal} {\bibinfo  {journal} {Phys. Rev. Lett.}\ }\textbf
  {\bibinfo {volume} {106}},\ \bibinfo {pages} {126802} (\bibinfo {year}
  {2011})}\BibitemShut {NoStop}%
\bibitem [{\citenamefont {Wong}\ \emph {et~al.}(2015)\citenamefont {Wong},
  \citenamefont {Wang}, \citenamefont {Jung}, \citenamefont {Pezzini},
  \citenamefont {DaSilva}, \citenamefont {Tsai}, \citenamefont {Jung},
  \citenamefont {Khajeh}, \citenamefont {Kim}, \citenamefont {Lee},
  \citenamefont {Kahn}, \citenamefont {Tollabimazraehno}, \citenamefont
  {Rasool}, \citenamefont {Watanabe}, \citenamefont {Taniguchi}, \citenamefont
  {Zettl}, \citenamefont {Adam}, \citenamefont {MacDonald},\ and\ \citenamefont
  {Crommie}}]{Crommie2015}%
  \BibitemOpen
  \bibfield  {author} {\bibinfo {author} {\bibfnamefont {Dillon}\ \bibnamefont
  {Wong}}, \bibinfo {author} {\bibfnamefont {Yang}\ \bibnamefont {Wang}},
  \bibinfo {author} {\bibfnamefont {Jeil}\ \bibnamefont {Jung}}, \bibinfo
  {author} {\bibfnamefont {Sergio}\ \bibnamefont {Pezzini}}, \bibinfo {author}
  {\bibfnamefont {Ashley~M.}\ \bibnamefont {DaSilva}}, \bibinfo {author}
  {\bibfnamefont {Hsin-Zon}\ \bibnamefont {Tsai}}, \bibinfo {author}
  {\bibfnamefont {Han~Sae}\ \bibnamefont {Jung}}, \bibinfo {author}
  {\bibfnamefont {Ramin}\ \bibnamefont {Khajeh}}, \bibinfo {author}
  {\bibfnamefont {Youngkyou}\ \bibnamefont {Kim}}, \bibinfo {author}
  {\bibfnamefont {Juwon}\ \bibnamefont {Lee}}, \bibinfo {author} {\bibfnamefont
  {Salman}\ \bibnamefont {Kahn}}, \bibinfo {author} {\bibfnamefont {Sajjad}\
  \bibnamefont {Tollabimazraehno}}, \bibinfo {author} {\bibfnamefont {Haider}\
  \bibnamefont {Rasool}}, \bibinfo {author} {\bibfnamefont {Kenji}\
  \bibnamefont {Watanabe}}, \bibinfo {author} {\bibfnamefont {Takashi}\
  \bibnamefont {Taniguchi}}, \bibinfo {author} {\bibfnamefont {Alex}\
  \bibnamefont {Zettl}}, \bibinfo {author} {\bibfnamefont {Shaffique}\
  \bibnamefont {Adam}}, \bibinfo {author} {\bibfnamefont {Allan~H.}\
  \bibnamefont {MacDonald}}, \ and\ \bibinfo {author} {\bibfnamefont
  {Michael~F.}\ \bibnamefont {Crommie}},\ }\bibfield  {title} {\enquote
  {\bibinfo {title} {Local spectroscopy of moir{\'e}-induced electronic
  structure in gate-tunable twisted bilayer graphene},}\ }\href {\doibase
  10.1103/PhysRevB.92.155409} {\bibfield  {journal} {\bibinfo  {journal} {Phys.
  Rev. B}\ }\textbf {\bibinfo {volume} {92}},\ \bibinfo {pages} {155409}
  (\bibinfo {year} {2015})}\BibitemShut {NoStop}%
\bibitem [{\citenamefont {Kim}\ \emph {et~al.}(2017)\citenamefont {Kim},
  \citenamefont {DaSilva}, \citenamefont {Huang}, \citenamefont {Fallahazad},
  \citenamefont {Larentis}, \citenamefont {Taniguchi}, \citenamefont
  {Watanabe}, \citenamefont {LeRoy}, \citenamefont {MacDonald},\ and\
  \citenamefont {Tutuc}}]{Kim2017}%
  \BibitemOpen
  \bibfield  {author} {\bibinfo {author} {\bibfnamefont {Kyounghwan}\
  \bibnamefont {Kim}}, \bibinfo {author} {\bibfnamefont {Ashley}\ \bibnamefont
  {DaSilva}}, \bibinfo {author} {\bibfnamefont {Shengqiang}\ \bibnamefont
  {Huang}}, \bibinfo {author} {\bibfnamefont {Babak}\ \bibnamefont
  {Fallahazad}}, \bibinfo {author} {\bibfnamefont {Stefano}\ \bibnamefont
  {Larentis}}, \bibinfo {author} {\bibfnamefont {Takashi}\ \bibnamefont
  {Taniguchi}}, \bibinfo {author} {\bibfnamefont {Kenji}\ \bibnamefont
  {Watanabe}}, \bibinfo {author} {\bibfnamefont {Brian~J.}\ \bibnamefont
  {LeRoy}}, \bibinfo {author} {\bibfnamefont {Allan~H.}\ \bibnamefont
  {MacDonald}}, \ and\ \bibinfo {author} {\bibfnamefont {Emanuel}\ \bibnamefont
  {Tutuc}},\ }\bibfield  {title} {\enquote {\bibinfo {title} {Tunable moir{\'e}
  bands and strong correlations in small-twist-angle bilayer graphene},}\
  }\href {http://www.pnas.org/content/114/13/3364.abstract} {\bibfield
  {journal} {\bibinfo  {journal} {Proceedings of the National Academy of
  Sciences}\ }\textbf {\bibinfo {volume} {114}},\ \bibinfo {pages} {3364}
  (\bibinfo {year} {2017})}\BibitemShut {NoStop}%
\bibitem [{\citenamefont {Hahn}(2006)}]{ITC}%
  \BibitemOpen
  \bibinfo {editor} {\bibfnamefont {Theo}\ \bibnamefont {Hahn}},\ ed.,\
  \href@noop {} {\emph {\bibinfo {title} {International Tables for
  Crystallography}}},\ \bibinfo {edition} {5th}\ ed.,\ Vol.\ \bibinfo {volume}
  {A: Space-group symmetry}\ (\bibinfo  {publisher} {Springer},\ \bibinfo
  {year} {2006})\BibitemShut {NoStop}%
\bibitem [{\citenamefont {Haldane}(1988)}]{Haldane}%
  \BibitemOpen
  \bibfield  {author} {\bibinfo {author} {\bibfnamefont {F.~D.~M.}\
  \bibnamefont {Haldane}},\ }\bibfield  {title} {\enquote {\bibinfo {title}
  {Model for a quantum hall effect without landau levels: Condensed-matter
  realization of the "parity anomaly"},}\ }\href {\doibase
  10.1103/PhysRevLett.61.2015} {\bibfield  {journal} {\bibinfo  {journal}
  {Phys. Rev. Lett.}\ }\textbf {\bibinfo {volume} {61}},\ \bibinfo {pages}
  {2015--2018} (\bibinfo {year} {1988})}\BibitemShut {NoStop}%
\bibitem [{\citenamefont {Soluyanov}\ and\ \citenamefont
  {Vanderbilt}(2011)}]{Z2Wannier}%
  \BibitemOpen
  \bibfield  {author} {\bibinfo {author} {\bibfnamefont {Alexey~A.}\
  \bibnamefont {Soluyanov}}\ and\ \bibinfo {author} {\bibfnamefont {David}\
  \bibnamefont {Vanderbilt}},\ }\bibfield  {title} {\enquote {\bibinfo {title}
  {Wannier representation of ${\mathbb{z}}_{2}$ topological insulators},}\
  }\href {\doibase 10.1103/PhysRevB.83.035108} {\bibfield  {journal} {\bibinfo
  {journal} {Phys. Rev. B}\ }\textbf {\bibinfo {volume} {83}},\ \bibinfo
  {pages} {035108} (\bibinfo {year} {2011})}\BibitemShut {NoStop}%
\bibitem [{\citenamefont {{Po}}\ \emph {et~al.}(2017)\citenamefont {{Po}},
  \citenamefont {{Watanabe}},\ and\ \citenamefont {{Vishwanath}}}]{Fragile}%
  \BibitemOpen
  \bibfield  {author} {\bibinfo {author} {\bibfnamefont {H.~C.}\ \bibnamefont
  {{Po}}}, \bibinfo {author} {\bibfnamefont {H.}~\bibnamefont {{Watanabe}}}, \
  and\ \bibinfo {author} {\bibfnamefont {A.}~\bibnamefont {{Vishwanath}}},\
  }\bibfield  {title} {\enquote {\bibinfo {title} {{Fragile Topology and
  Wannier Obstructions}},}\ }\href@noop {} {\bibfield  {journal} {\bibinfo
  {journal} {ArXiv e-prints}\ } (\bibinfo {year} {2017})},\ \Eprint
  {http://arxiv.org/abs/1709.06551} {arXiv:1709.06551} \BibitemShut {NoStop}%
\end{thebibliography}%

\clearpage
\appendix
\section{Conjugate small angles
\label{sec:conj}}
In this appendix, we discuss the ``conjugate'' structures defined by angles characterized by $\theta(m,r) = \pi/3 - \phi$ in Eq.\ \eqref{eq: commensurate angles} for some $0<\phi\ll1$. As discussed in the main text, physically these structures are also small-angle TBG, and so are as relevant as the other cases of $\theta \ll 1$ typically discussed in the literature.

First, we note that the conjugate pair of angles $\theta(m,r)$ and $\pi/3 - \theta(m,r)$ have the same lattice constant. To this end, check that, for coprime positive integers $m,r$, we have
\beq \begin{split}
\theta (r,3m) =& \frac{\pi}{3} - \theta(m,r);\\
 L\left(\frac{r}{ {\rm gcd}(r,3)},\frac{3m}{ {\rm gcd}(r,3)} \right) =& L(m,r).
\end{split}
\eeq
Therefore, when one says that the {\it physical} twist angle is a commensurate angle $\theta(m,1)$ for some $m$, and that the {\it exact} lattice constant is $L'(\theta)$, a priori one does not know if it corresponds to the case of $\theta(m,1)$ or $\theta(1,3m)$ in the parameterization of Eq.\ \eqref{eq: commensurate angles}.

The astute readers might have noticed a conundrum: while we have argued that $\theta(m,1)$ and $\theta(1,3m)$ are conjugate angles that can correspond to the same physical system, in Fig.\ \ref{fig:folding} they belong respectively to types I and II structures and have different BZ folding patterns.
For small twist angles, this conundrum can be readily resolved by focusing on the type II case of $ \theta = \pi/3 - \phi$, where $ \phi\ll1$ is a type I angle.
While Fig.\ \ref{fig:folding} suggests that K and K$_{\theta}$ are now folded to the same \moire K point, caution must be used when the angle is really $\theta = \pi/3 - \phi$ with $\phi \ll 1$. For such structures, $|\vec K_{\theta} - \vec K| \simeq |\vec K_{\pi/3} - \vec K| \sim \mathcal O(1/a)$, but $|\vec K_{\theta} - \vec K'| \sim \mathcal O(\phi/a)$.
Therefore, the correct assignment in this case is to group the K' and K$_\theta$ points into one valley, and again the symmetry-allowed coupling between K and K$_{\theta}$ is suppressed by the approximate valley charge conservation. This leads to the same physical definition of valleys as in the type I case with twist angle $\phi$.

Finally, let us clarify on the symmetry representations for the conjugate type II structures. If one chooses the twisting center to be an aligned hexagon center, the point-group origin will again coincide with the twisting center and the analysis follows that discussed in the main text. The more nontrivial case is when one chooses to twist about a common carbon site, as is done in many existing works, say Refs.\ \onlinecite{Mele2010, NoahLiang, Oskar}. In this case, a conjugate type II twist of $\theta =  \pi/3 - \phi$ with $\phi\ll 1$ will place the twisting center into a locally AB region. Furthermore, as there will be an aligned hexagon center elsewhere, the proper choice of point-group origin will always be different from the twisting center.

Physically, such a lattice is essentially identical to one built from twisting about an aligned hexagon center, and therefore we would seek to reconcile the analysis with that in Eq.\ \eqref{eq:D6Rep} the main text.
There are two main changes compared to the type I analysis there. First, the momentum mapping becomes different, with  K$_{\theta}$ and K landing on K$_{\rm m}$ (Fig.\ \ref{fig:folding}); second, the rotation symmetry about the twisting center becomes $C_{3}^{\rm AB} = C_{3}^{\rm C}$. Again, by combining Eqs.\ \eqref{eq:C3Crep} and \eqref{eq:KExpand} in the main text, we readily conclude that the representations for $C_3^{\rm AB}$ are $( 1,\omega) \cup (1, \omega)$.
To reconcile with Eq.\ \eqref{eq:D6Rep}, it remains to notice that $C_{3}^{\rm AB} = T_{\vec t_1}^{-1} C_{3}^{\rm AA}$ (Fig. \ref{fig:commensurate}). As $T_{\vec t_1}^{-1} | \psi_{\vec K_{\rm m}} ^{\sigma}, \pm \rangle = | \psi_{\vec K_{\rm m}} ^{\sigma}, \pm \rangle  \omega$, we find that the representation for $C_3^{\rm AA}$ is $\omega \times \left( ( 1,\omega) \cup (1, \omega) \right) = ( \omega,\omega^*) \cup (\omega,\omega^*)$, as one expects.

\section{Mirror-eigenvalue obstruction and chirality-obstruction} \label{app: relation_obstruction}

In this appendix we present the detailed arguments that relate the mirror-eigenvalue obstruction and the chirality-obstruction, as stated in Sec. \ref{sec: relation_obstruction}.

To start, let us first record the symmetry algebra:
\beq \label{eq: symmetry algebra}
M_y^2=1,
\quad
(C_2\mc{T})^2=1,
\quad
M_y(C_2\mc{T})=(C_2\mc{T})M_y
\eeq
In the following, we will show that having mirror eigenvalues $\pm 1$ at the $M$ point is equivalent to having the same chirality for the two Dirac points at $K$ and $K'$. To show this, we take three steps:

\begin{itemize}
	
	\item[] 1. On an {\em open} region of the mBZ that covers the $K$, $K'$ and $M$ points, the action of $C_2\mc{T}$ can always be chosen to be
	\beq \label{eq: canonical C2T 00}
	\psi(\vec k)\rightarrow\sigma_x\mc{K}\psi(\vec k)
	\eeq
	for all $k$ in this region, where $\mc{K}$ denotes complex conjugation. This choice can be made while having a smooth basis.
	
	\item[] 2. When the above choice of the $C_2\mc{T}$ action is made, if the mirror eigenvalues at $M$ are $\pm 1$, we can choose the action of $M_y$ to be
	\beq \label{eq: canonical My 1}
	\psi(\vec k)\rightarrow\sigma_x\psi(\vec k')
	\eeq
	for all $k$ in this region, where $k'$ is the $M_y$ partner of $k$. If the mirror eigenvalues at $M$ are the same, then we can choose the action of $M_y$ to be
	\beq \label{eq: canonical My 2}
	\psi(\vec k)\rightarrow\eta_M\psi(\vec k')
	\eeq
	with $\eta_M=\pm 1$. For either case, the above choice can be made while having the basis smooth.
	
	\item[] 3. The above two symmetry actions guarantee that the chiralities of the two Dirac points at $K$ and $K'$ are the same (opposite) if the mirror eigenvalues at $M$ are opposite (same).
	
\end{itemize}

Below we prove these statements one by one.

\subsection{Action of $C_2\mc{T}$}

First assume a generic action of $C_2\mc{T}$ under a smooth basis:
\beq \label{eq: original C2T}
\psi(\vec k)\rightarrow U(\vec k)\psi(\vec k)
\eeq
where $U(\vec k)$ is a 2$\times$2 unitary matrix that satisfies $U^*U=1$, because $(C_2\mc{T})^2=1$. The generic form of such $U(\vec k)$ is
\beq \label{eq: generic C2T rep}
U(\vec k)=e^{i\theta_0(\vec k)}\left(a_0(\vec k)+ia_1(\vec k)\sigma_x+ia_3(\vec k)\sigma_z\right)
\eeq
The meaning of a smooth basis is that $a_{0,1,2,3}(\vec k)$ and $\theta_0(\vec k)$ are smooth functions of $\vec k$ in this open region. Notice due to the lack of a term proportional to $\sigma_y$ in $U(\vec k)$, $U(\vec k)$ can be diagonalized by an {\em orthogonal} transformation. Furthermore, the orthogonal matrix corresponding to this transformation can be chosen to be a smooth function of $\vec k$ because $U(\vec k)$ is smooth.

In order to find a basis in which the action of $C_2\mc{T}$ is given by (\ref{eq: canonical C2T 00}), we need to find a unitary $V(\vec k)$ such that
\beq \label{eq: canonical C2T 0}
V(\vec k)\psi(\vec k)\rightarrow \sigma_x\mc{K}V(\vec k)\psi(\vec k)
\eeq
under $C_2\mc{T}$. If this unitary can be found, then the action of $C_2\mc{T}$ is given by (\ref{eq: canonical C2T 00}) on the basis $V(\vec k)\psi(\vec k)$. It is not hard to see that this is equivalent to finding a unitary $\tilde V(k)$, such that under $C_2\mc{T}$
\beq \label{eq: canonical C2T}
\tilde V(\vec k)\psi(\vec k)\rightarrow \mc{K}\tilde V(\vec k)\psi(\vec k)
\eeq
So below we will show that this latter $\tilde V(\vec k)$ exists.

Combining (\ref{eq: original C2T}) and (\ref{eq: canonical C2T}) yields
\beq
U(\vec k)=\tilde V(\vec k)^T\tilde V(\vec k)
\eeq
Because $U(\vec k)$ can be diagonalized by an orthogonal transformation, there must be a solution of $\tilde V(\vec k)$ to the above equation, and the solution can be made smooth as a function of $\vec k$, because $U(k)$ is a smooth function of $\vec k$ itself.

This means that, within this open region, the action of $C_2\mc{T}$ can always be chosen as
\beq
\psi(\vec k)\rightarrow\sigma_x\mc{K}\psi(\vec k)
\eeq
while preserving the smoothness of the basis. This concludes the first step listed above.

\subsection{Action of $M_y$}

Now we go to the smooth basis under which the action of $C_2\mc{T}$ is given by (\ref{eq: canonical C2T 00}), and assume a generic $M_y$ action under this basis
\beq
\psi(\vec k)\rightarrow M(\vec k)\psi(\vec k')
\eeq
with a unitary $M(\vec k)$. The symmetry algebra (\ref{eq: symmetry algebra}) implies that
\beq
\begin{split}
M(\vec k)M(\vec k')=&M(\vec k')M(\vec k)=1,\\
M(\vec k)^*=& \sigma_xM(\vec k)\sigma_x
\end{split}
\eeq
This means there are two possible types of $M(\vec k)$:
\beq \label{eq: first-type-M}
M(\vec k)=\eta(\vec k)e^{i\sigma_z\theta(\vec k)}
\eeq
where $\eta(\vec k)=\pm 1$, or
\beq \label{eq: second-type-M}
M(\vec k)=\cos\theta(\vec k)\sigma_x+\sin\theta(\vec k)\sigma_y
\eeq
Furthermore, the type of the mirror actions at $k$ and $k'$ must be the same. If both at $k$ and $k'$ the mirror action is of the first type, then
\beq
\eta(\vec k)e^{i\sigma_z\theta(\vec k)}\eta(\vec k')e^{i\sigma_z\theta(\vec k')}=1
\eeq
If both at $k$ and $k'$ the mirror action is of the second type, then $\theta(k)=\theta(k')$.

Now we consider the case where the mirror eigenvalue at $M$ is $\pm 1$. Then the mirror action at $M$ must be of the second type, because the first type will not give rise to two different eigenvalues at $M$. Without loss of generality, we can take the action of $M_y$ at $M$ to be
\beq
\psi(\vec k=M)\rightarrow\sigma_x\psi(\vec k=M)
\eeq

Next we look for a unitary $W(\vec k)$ such that under $M_y$
\beq
W(\vec k)\psi(\vec k)\rightarrow \sigma_xW(\vec k')\psi(\vec k')
\eeq
If such a $W(\vec k)$ can be found, the action of $M_y$ can be chosen to be (\ref{eq: canonical My 1}). This requires that
\beq \label{eq: requirement-1}
W(\vec k)M(\vec k)=\sigma_xW(\vec k')
\eeq
Notice this requirement automatically implies $W(\vec k')M(\vec k')=\sigma_xW(\vec k)$ by noting that $M(\vec k)M(\vec k')=1$. We need to choose $W(\vec k)$ such that the $C_2\mc{T}$ action is still given by (\ref{eq: canonical C2T 00}), which requires
\beq \label{eq: requirement-2}
W(\vec k)^*=\sigma_xW(\vec k)\sigma_x
\eeq

The smoothness of the basis means that in this region $M(k)$ is always of the second type, given by (\ref{eq: second-type-M}). Then to satisfy the requirements (\ref{eq: requirement-1}) and (\ref{eq: requirement-2}), we can choose
\beq
W(\vec k)=W(\vec k')=e^{-i\frac{\sigma_z}{2}\theta(\vec k)}
\eeq
This is indeed a smooth transformation given that $\theta(\vec k)=\theta(\vec k')$ in this case, as discussed earlier.

This tells us that as long as the mirror eigenvalues at $M$ are $\pm 1$, we can always choose the mirror action in this region to be given by (\ref{eq: canonical My 1}), while preserving the smoothness of the basis.

In contrast, if both mirror eigenvalues at $M$ are the same, the mirror action at $M$ must be of the first type. That is, we can write the action of $M_y$ at $M$ as
\beq
\psi(\vec k=M)\rightarrow\eta_M\psi(\vec k=M)
\eeq
where $\eta_M=\pm 1$ is the mirror eigenvalue at $M$. The smoothness of the basis implies that $M(k)$ in the entire open region is of the first type, given by (\ref{eq: first-type-M}). In addition, in this entire open region $\eta(\vec k)=\eta_M$. Then we can find a unitary $W(\vec k)$ transformation, such that
\beq
W(\vec k)M(\vec k)=\eta_MW(\vec k')
\eeq
so that the mirror action is given by (\ref{eq: canonical My 2}) in the entire open region, and the basis is still smooth while the action of $C_2\mc{T}$ is still given by (\ref{eq: canonical C2T 00}). More explicitly, we can choose
\beq
W(\vec k)=W(\vec k')^\dag=e^{-i\frac{\sigma_z}{2}\theta(\vec k)}
\eeq
This concludes the second step listed above.

\subsection{Relative chiralities}

Now we go to a basis where the action of $C_2\mc{T}$ is given by (\ref{eq: canonical C2T 00}), and the action of $M_y$ is given by either (\ref{eq: canonical My 1}) or (\ref{eq: canonical My 2}), depending on whether the mirror eigenvalues at $M$ are different or identical. This $C_2\mc{T}$ action constrains the first-quantized Hamiltonian to be
\beq
H(k)=n_0(\vec k)+n_1(\vec k)\sigma_x+n_2(\vec k)\sigma_y
\eeq
And the winding of $(n_1(\vec k), n_2(\vec k))^T$ along a closed path defines the chirality along this closed path. From this definition, we see that only gapless points in the Brillouin zone contribute to the net chirality.

If the mirror eigenvalues are opposite at $M$, that is, the action of $M_y$ is given by (\ref{eq: canonical My 1}), then $n_1(\vec k)=n_1(\vec k')$ and $n_2(\vec k)=-n_2(\vec k')$, where $\vec k'$ is the $M_y$-partner of $\vec k$. Now consider a pair of gapless points in the Brillouin zone that are related by $M_y$, and draw a small closed loop around each of them. It is straightforward to see that the windings around these two loops are the same. In contrast, if the mirror eigenvalues are identical at $M$, that is, the action of $M_y$ is given by (\ref{eq: canonical My 2}), then $n_{1,2}(\vec k)=n_{1,2}(\vec k')$. It is straightforward to see that the windings around a pair of gapless points related by $M_y$ are opposite. This concludes that third step listed above.

In short, when $M_y$ is preserved, having opposite (same) mirror eigenvalues at $M$ is equivalent to having same (opposite) chiralities at a pair of gapless points related by $M_y$. Given that both $K$ and $K'$ are host Dirac points and they are related by $M_y$, the net chirality of the entire Brillouin zone is nonzero (zero) if the mirror eigenvalues at $M$ are opposite (identical). Upon breaking $M_y$, the net chirality cannot change due to its discrete nature. This enables us to check the net chirality by simply looking at the eigenvalues of $M_y$ at its high symmetry points.

\end{document}